\documentclass[iop, tighten, numberedappendix]{emulateapj}

\usepackage{amsmath}
\usepackage[usenames, dvipsnames]{color}
\usepackage{hyperref}
\hypersetup{colorlinks=True, citecolor=blue, linkcolor=Purple}

\newcommand{\Msun}{\mbox{$\rm M_{\odot}$}}
\newcommand\av{{$A_V$}}
\newcommand\dav{{$dA_V$}}
\newcommand{\sfr}{\:{\Msun\:${\rm yr}^{-1}$}}
\newcommand{\sfrd}{\:{\Msun\:${\rm yr}^{-1}\:{\rm kpc}^{-2}$}}
\newcommand{\lstar}{L$_\star$~}
\newcommand{\match}{\texttt{MATCH}}
\newcommand{\hst}{\textit{HST}}
\newcommand{\sigsfr}{$\Sigma_{\rm SFR}$}

\def\about  {\hbox{$\sim$}}

\slugcomment{Accepted April 8, 2015}

\begin{document}

\shortauthors{Lewis et al.}
\shorttitle{}

\title{The Panchromatic Hubble Andromeda Treasury XI: The Spatially-Resolved Recent Star Formation History of M31\altaffilmark{*}}

\altaffiltext{*}{Based on observations made with the NASA/ESA Hubble Space Telescope, obtained at the Space Telescope Science Institute, which is operated by the Association of Universities for Research in Astronomy, Inc., under NASA contract NAS 5-26555. These observations are associated with program \#12055.}

\author{Alexia~R.~Lewis\altaffilmark{1},
Andrew~E.~Dolphin\altaffilmark{2},
Julianne~J.~Dalcanton\altaffilmark{1},
Daniel~R.~Weisz\altaffilmark{1,12},
Benjamin~F.~Williams\altaffilmark{1}, 
Eric F. Bell\altaffilmark{3},
Anil C. Seth\altaffilmark{4},
Jacob E. Simones\altaffilmark{5},
Evan D. Skillman\altaffilmark{5},
Yumi Choi\altaffilmark{1},
Morgan Fouesneau\altaffilmark{6},
Puragra Guhathakurta\altaffilmark{7},
Lent~C.~Johnson\altaffilmark{1},
Jason S. Kalirai\altaffilmark{8},
Adam K. Leroy\altaffilmark{9},
Antonela Monachesi\altaffilmark{10},
Hans-Walter~Rix\altaffilmark{6},
Andreas~Schruba\altaffilmark{11}
}

\altaffiltext{1}{Department of Astronomy, University of Washington, Box 351580, Seattle, WA 98195, USA;\\arlewis@astro.washington.edu}
\altaffiltext{2}{Raytheon, 1151 E. Hermans Road, Tucson, AZ 85756, USA}
\altaffiltext{3}{Department of Astronomy, University of Michigan, 1085 S. University Ave., Ann Arbor, MI 48109, USA}
\altaffiltext{4}{Department of Physics and Astronomy, University of Utah, Salt Lake City, UT 84112, USA}
\altaffiltext{5}{Minnesota Institute for Astrophysics, University of Minnesota, 116 Church Street SE, Minneapolis, MN 55455, USA}
\altaffiltext{6}{Max-Planck-Institut f\"ur Astronomie, K\"onigstul 17, D-69117 Heidelberg, Germany}
\altaffiltext{7}{UCO/Lick Observatory, University of California at Santa Cruz, 1156 High Street, Santa Cruz, CA 95064, USA}
\altaffiltext{8}{Space Telescope Science Institute, 3700 San Martin Drive, Baltimore, MD 21218, USA}
\altaffiltext{9}{National Radio Astronomy Observatory, 520 Edgemont Road, Charlottesville, VA 22903, USA}
\altaffiltext{10}{Max-Planck-Institut f\"ur Astrophysik, Karl-Schwarzschild-Str. 1, D-85748 Garching, Germany}
\altaffiltext{11}{Max-Planck-Institut f\"ur extraterrestrische Physik, Giessenbachstrasse 1, 85748 Garching, Germany}
\altaffiltext{12}{Hubble Fellow}
\altaffiltext{}{}

\begin{abstract}
We measure the recent star formation history (SFH) across M31 using optical images taken with the \textit{Hubble Space Telescope} as part of the Panchromatic Hubble Andromeda Treasury (PHAT). We fit the color-magnitude diagrams in \about9000 regions that are \about100 pc $\times$ 100 pc in projected size, covering a 0.5 square degree area (\about380 kpc$^2$, deprojected) in the NE quadrant of M31. We show that the SFHs vary significantly on these small spatial scales but that there are also coherent galaxy-wide fluctuations in the SFH back to \about500 Myr, most notably in M31's 10-kpc star-forming ring. We find that the 10-kpc ring is at least 400 Myr old, showing ongoing star formation over the past \about500 Myr. This indicates the presence of molecular gas in the ring over at least 2 dynamical times at this radius. We also find that the ring's position is constant throughout this time, and is stationary at the level of 1 km s$^{-1}$, although there is evidence for broadening of the ring due to diffusion of stars into the disk. Based on existing models of M31's ring features, the lack of evolution in the ring's position makes a collisional ring origin highly unlikely.  Besides the well-known 10-kpc ring, we observe two other ring-like features. There is an outer ring structure at 15 kpc with concentrated star formation starting \about80 Myr ago. The inner ring structure at 5 kpc has a much lower star formation rate (SFR) and therefore lower contrast against the underlying stellar disk. It was most clearly defined \about200 Myr ago, but is much more diffuse today. We find that the global SFR has been fairly constant over the last \about500 Myr, though it does show a small increase at 50 Myr that is 1.3 times the average SFR over the past 100 Myr. During the last \about500 Myr, \about60\% of all SF occurs in the 10-kpc ring. Finally, we find that in the past 100 Myr, the average SFR over the PHAT survey area is $0.28\pm0.03$\sfr\ with an average deprojected intensity of $7.3\times10^{-4}$\sfrd, which yields a total SFR of \about0.7\sfr\ when extrapolated to the entire area of M31's disk. This SFR is consistent with measurements from broadband estimates. 
\end{abstract}


\section{Introduction}
\label{sec:intro}

A galaxy's star formation history (SFH) encodes much of the physics controlling its evolution. It tells us about the evolution of the star formation rate (SFR) throughout the galaxy, the evolution of the mass and metallicity distributions, and the movement of stars within the galaxy. In addition to the global evolution, focusing on the recent SFH ($<$1 Gyr) reveals the relationships between stars and the gas and dust from which they form and can be used to constrain models of star formation (SF) propagation and/or dissolution. For this type of study to be possible, however, we first need a spatially-resolved view of the past SFH with sufficient resolution to probe the relevant physical scales.

Broad SFH constraints can be derived by looking at the properties of the galaxy population across cosmic time. However, examining the integrated properties of distant galaxies provides limited information on how individual galaxies form and evolve. While such integrated light studies benefit from large sample sizes, the final results are limited to conclusions about the SFHs of general galaxy types (e.g., based on bins of mass, luminosity, or color), and cannot say anything definitive about the physics that controls the evolution of individual galaxies.

To appropriately examine the evolution of individual galaxies, it is necessary to study well-resolved nearby galaxies for which there exists large amounts of ancillary data. With such data, one can, for example, analyze the relationship between star SF and gas in the spatially-resolved Kennicutt-Schmidt law \citep[e.g.,][]{Kennicutt2007a, Bigiel2008a}, understand the evolution of a galaxy's gas reservoir \citep[e.g.,][]{Leroy2008a, Schruba2010a, Bigiel2011a, Leroy2013b}, and calibrate SFR indicators \citep[e.g.,][]{Calzetti2007a, Li2013a}, among many others. However, these studies have historically been restricted to using only the current SFR where `current' is the average over some timescale characteristic of a given SFR indicator. These studies, therefore, cannot probe the evolution of these relationships with time or on small physical scales where the SFR indicators break down \citep[e.g.,][]{Leroy2012a}.

For a more detailed analysis of the recent SFH, resolved stellar populations are the gold standard. Using individual stars, we can examine the evolution of a galaxy archaeologically by analyzing the color-magnitude diagram (CMD) as a function of position within the galaxy. Embedded within the CMD is the history of SF and metallicity evolution of the galaxy. Although recovering this information is not completely assumption-free (we must make choices about the initial mass function (IMF), stellar models, constancy of SFR within time bins, etc), it is the only way to make a time-resolved measurement of the SFR. It also has the ability to recover the SFR on much finer physical scales.

CMD fitting has most often been used to probe low mass galaxies \citep[e.g.,][]{Gallart1999a, Harris2004a, Cole2007a, Harris2009a, Monelli2010a, Weisz2011a, Monachesi2012a, Weisz2014a} because they are the most numerous type of galaxy in the Local Volume, which is one of the few places where galaxies can be sufficiently well resolved. The technique has been used to examine a few individual larger galaxies \citep[e.g.,][]{Wyder1998a, Hernandez2000a, Bertelli2001a, Williams2002a, Williams2003a, Brown2006a, Brown2007a, Brown2008a, Williams2009a, Williams2010a, Gogarten2010a, Bernard2012a, Bernard2015a}, but these studies have been limited to either small fields spread across the disk and/or halo or have low spatial resolution such that it is difficult to pick out detailed features present in the galaxy. This technique has never been used to contiguously and uniformly recover the recent SFH of an \lstar galaxy with high resolution. 

In this paper, we present the first finely spatially-resolved recent SFH of a significant part of the \lstar galaxy, Andromeda (M31). M31 has been the target of many photometric \citep[e.g.,][]{Brown2006a, Barmby2006a, Gordon2006a, Dalcanton2012a, Bernard2012a, Ford2013a, Sick2014a} and spectroscopic \citep[e.g.,][]{Ibata2004a, Guhathakurta2006a, Kalirai2006a, Koch2008a, Gilbert2009a, Dorman2012a, Gilbert2014a} studies due to its proximity and similarity to the Milky Way. M31 is the ideal place to examine processes in \lstar galaxies; it is close enough to be well resolved into stars with the \textit{Hubble Space Telescope} (\hst) but does not not face the same obstacles as studies in the Milky Way, which are plagued by uncertainties due to line-of-sight reddening and challenging distance measurements.

We generate maps of the recent ($<$0.5 Gyr) SFH in M31 using resolved stars from recent observations of M31 taken as part of the Panchromatic Hubble Andromeda Treasury \citep[PHAT;][]{Dalcanton2012a}. While other studies have examined the SFH in M31 using resolved stars \citep{Williams2003a, Brown2006a, Brown2007a, Brown2008a, Davidge2012a, Bernard2012a}, none have been as finely resolved as the work we present here.

With the resulting spatially- and temporally-resolved recent SFHs, we can see where stars form within the galaxy and how that SF evolves across the galaxy, whether it's a single star-forming event or propagation across the disk. The maps also provide clues about the evolution of spatial structure on a variety of different scales; while we know a great deal about small-scale SF within molecular clouds and large-scale SF within the galactic environment, the maps we derive bridge these two scales. Recent SFHs also enable the analysis of fluctuations in the recent SFR. This is especially significant for SF relations, such as the Kennicutt-Schmidt relation \citep[][]{Schmidt1959a, Kennicutt1989a}, which often assumes a constant SFR over the timescale of the tracer used. While this paper deals only with the SFHs themselves, it is the first in a series of papers on the SF, dust, and ISM contents of M31 on small spatial scales (Lewis et al., in prep).

This paper is organized as follows: We describe the data used in Section \ref{sec:data}. In Section \ref{sec:SFHderive}, we explain the method by which we recover the SFHs in each region. We present the resulting SFH maps in Section \ref{sec:results} and discuss features of the maps in Section \ref{sec:discussion}. We summarize the results in Section \ref{sec:conclusion}.

\section{PHAT Data}
\label{sec:data}

\begin{figure}[]
\centering
\includegraphics[width=\columnwidth]{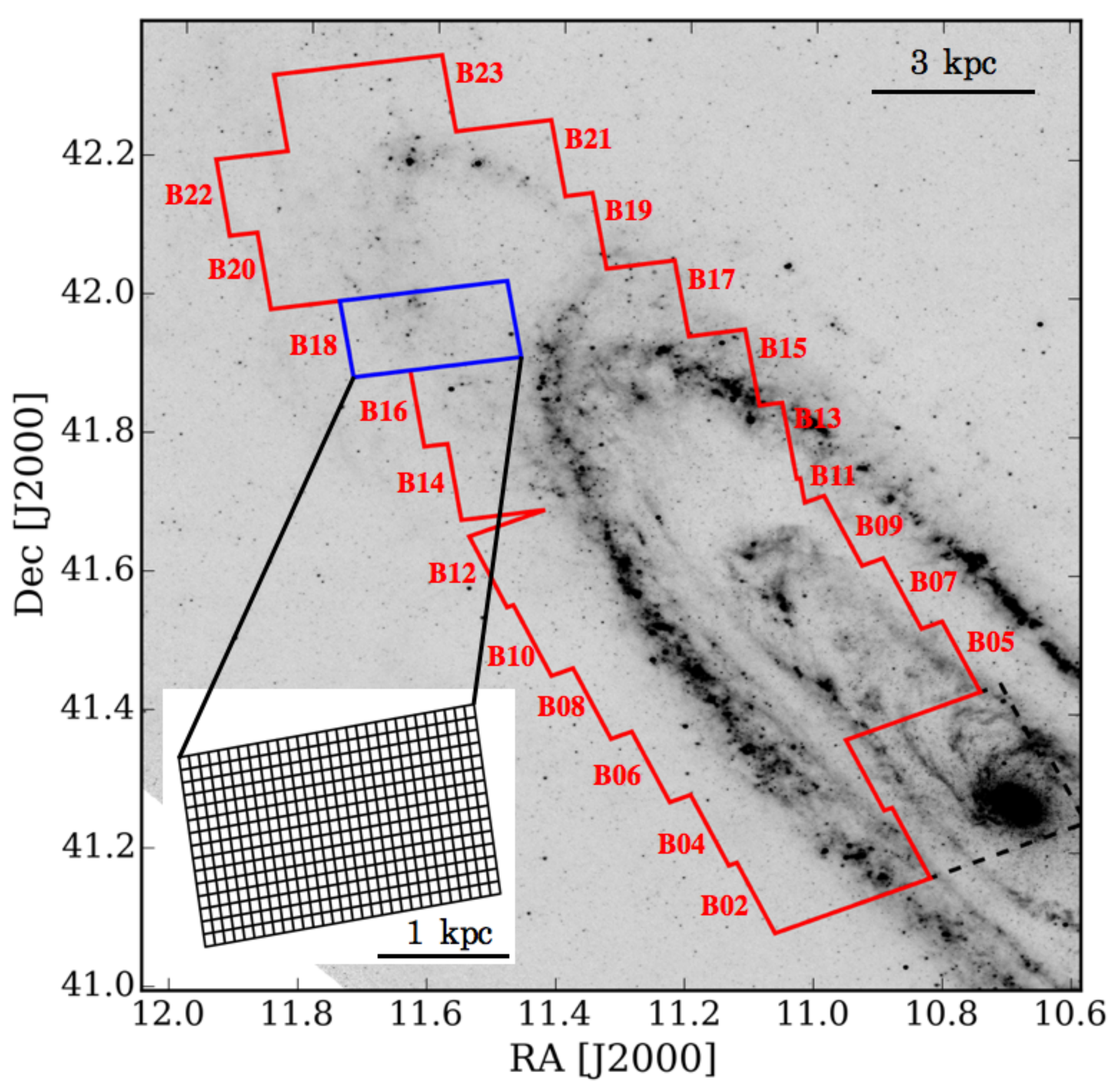}
\caption{The PHAT NIR footprint overlaid on a 24 \micron\ image \citep{Gordon2006a} of M31. The thick red line denotes the area for which we compute SFHs in this paper. We have left out the regions closest to the bulge along the major axis (dashed black line) because crowding errors are large and CMD depth is shallow, which makes fitting a reliable SFH difficult. The inset shows an example of our binning scheme, shown here for Brick 18, which is outlined in blue. We have included scale bars for the large image as well as for the inset. The bricks are labeled with their numbers just exterior to the brick outline. We use these brick numbers throughout the text. The image is oriented such that north is up and east is to the left.} 
\label{fig:phat_area}
\end{figure}

We derive the spatially resolved SFHs using photometry from the PHAT survey. PHAT surveyed the northeast quadrant of M31 in six filters, from the near-UV to the near-IR, measuring the properties of \about117 million stars. Full details of the survey can be found in \citet{Dalcanton2012a} and the photometry is described in \citet{Williams2014a}. Figure \ref{fig:phat_area} shows a 24 \micron\ image \citep{Gordon2006a} of M31 with the PHAT footprint overlaid. In this paper, we examine the SFH inside the solid red region; we have excluded the region closest to the bulge (black dashed line)  where crowding errors are large and the depth of the CMD is shallow, making reliable CMD fitting difficult.

\subsection{Photometry and Creation of Single-Brick Catalogs}
\label{subsec:catalog}
We use optical photometry (F475W and F814W filters) from the \texttt{.gst} catalogs, which were compiled following procedures developed by \citet{Dolphin2000a} as described in \citet{Dalcanton2012a}. The stars in this catalog have S/N $>$ 4 in both filters and pass stringent goodness-of-fit cuts. These cuts leave stars with the highest quality photometry but with higher incompleteness in more crowded regions. The incompleteness is worse in the inner galaxy (inside \about3 kpc), which we exclude from our analysis, and also affects the centers of stellar clusters. This latter limitation is not a problem for this paper because clusters contain only a few percent of the recent SF (Johnson et al., in prep.). Moreover, we are interested in the SFHs of the field stars, and picking out stars in the centers of dense stellar clusters is not necessary.

The survey was split into 23 regions called `bricks'; each brick is \about1.5 kpc $\times$ 3 kpc in projected size. Odd-numbered bricks extend from the galactic center to the outer disk along the major axis. Even-numbered bricks sit adjacent to them at larger radii along the minor axis. Each brick is subdivided into 18 `fields', each with a (projected) size of \about500 pc $\times$ 500 pc. In the optical, adjacent fields overlap to cover the ACS chip gap. The SFHs presented in this paper are derived on a brick-by-brick basis. To create a single brick catalog from the 18 `field' catalogs, we use the smaller IR brick footprint divided into 18 non-overlapping regions which roughly describe the IR field footprints. In each of these fields, we select all of the stars in the corresponding optical catalog that fall within the IR field boundaries. We fill in the chip gap using two adjacent fields, selecting only the stars that fall within the desired portion of the chip gap. Three fields in Brick 11 were not observed; this area is completely covered by overlap from Brick 09 so that there is continuous coverage over the survey area.

The result of this process is the creation of a single brick catalog of all stars detected in the optical filters that fall within the IR footprint, filling in the chip gap and eliminating duplication of stars in the overlap regions. We then grid each brick into 450 approximately equal sized, non-overlapping regions that are \about100 pc $\times$ 100 pc in projected size for a total of \about9000 regions across the survey area. In Figure \ref{fig:phat_area}, B18 is outlined in blue. The inset shows the binning scheme used within each brick.


\subsection{Artificial Star Tests}
\label{subsec:asts}

Even with the resolution of \hst, crowding in regions of high stellar density can strongly affect the photometry. Many faint stars cannot be resolved in the dense field of brighter stars. In addition, faint stars, that would otherwise not be detected, are biased brighter by blending with neighboring stars. This also affects brighter stars, but to a lesser degree.

To characterize photometric completeness and to account for the observational errors that result from crowding, we perform extensive artificial star tests (ASTs).  Briefly, we insert fake stars into each image and run the photometry as normal. We then test for recovery of these fake stars and measure the difference between the input and recovered magnitude if a star was detected. We adopt the magnitude at which 50\% of the stars are recovered as our limiting magnitude when solving for the SFHs. The completeness limits used for each brick are given in Table \ref{tab:completeness}.We refer the reader to \citet{Dalcanton2012a} for further details. 

We inserted \about100,000 artificial stars individually into each ACS field-of-view. We combined the resulting ASTs into brick-wide catalogs in the same way as the photometry. When running the SFHs for a given 100 pc region, we select the fake stars from a 5$\times$5 grid of adjacent regions, such that the ASTs come from a 500$\times$500 pc$^2$ region centered on the region of interest. Each of these larger regions contains the results of \about50,000 ASTs.

\begin{deluxetable}{ccccc}
\tabletypesize{\footnotesize}
\tablecaption{50\% Completeness Limits and Isochrone Shifts}
\tablecolumns{5}
\tablewidth{0pt}
\tablehead{
  \colhead{Brick} & 
  \colhead{m$_{F475W}$} & 
  \colhead{m$_{F814W}$} &
  \colhead{$\sigma_{\log \mathrm{T}_{\mathrm{eff}}}$} & 
  \colhead{$\sigma_{\mathrm{M}_{\mathrm{bol}}}$} \\ 
  \colhead{Number} & 
  \colhead{(mag)} & 
  \colhead{(mag)} &
  \colhead{} & 
  \colhead{}
}
\startdata
02 & 27.2 & 26.1 & 0.019 & 0.22 \\
04 & 27.1 & 25.9 & 0.019 & 0.21 \\
05 & 26.4 & 25.0 & 0.019 & 0.18 \\
06 & 27.2 & 26.0 & 0.019 & 0.22 \\
07 & 26.8 & 25.4 & 0.019 & 0.18 \\
08 & 27.2 & 26.0 & 0.019 & 0.22 \\
09 & 27.1 & 25.8 & 0.019 & 0.21 \\
10 & 27.2 & 26.1 & 0.019 & 0.23 \\
11 & 27.1 & 25.9 & 0.019 & 0.21 \\
12 & 27.3 & 26.2 & 0.019 & 0.24 \\
13 & 27.2 & 26.1 & 0.019 & 0.23 \\
14 & 27.3 & 26.2 & 0.019 & 0.24 \\
15 & 27.4 & 26.2 & 0.019 & 0.25 \\
16 & 27.4 & 26.4 & 0.019 & 0.26 \\
17 & 27.5 & 26.4 & 0.020 & 0.26 \\
18 & 27.8 & 26.7 & 0.020 & 0.29 \\
19 & 27.6 & 26.7 & 0.020 & 0.28 \\
20 & 27.8 & 26.9 & 0.020 & 0.30 \\
21 & 27.8 & 26.9 & 0.020 & 0.30 \\
22 & 27.8 & 26.9 & 0.020 & 0.30 \\
23 & 27.8 & 26.9 & 0.020 & 0.30 
\enddata
\label{tab:completeness}
\tablecomments{Column 1 contains the brick number. Columns 2 and 3 list the 50\% completeness limits in F475W and F814W, respectively. Columns 4 and 5 contain the shifts in $\log$ T$_{\mathrm{eff}}$ and M$_{bol}$ used when computing the systematic uncertainties.}
\end{deluxetable}

\section{Derivation of the Star Formation Histories}
\label{sec:SFHderive}

We derive SFHs using only the optical data from the F475W and F814W filters. These filters provide the deepest CMDs and the greatest leverage for the recent SFHs of interest in this paper. A more detailed discussion of our filter choice can be found in Appendix \ref{app:filterchoice}.

\subsection{Fitting the Star Formation History}
\label{subsec:SFH_fit}

We derive the SFHs using the CMD fitting code \match\ described in \citet{Dolphin2002a}. The user specifies desired ranges in age, metallicity, distance, and extinction. The code also requires a choice of IMF and a binary fraction. It then populates CMDs at each combination of age and metallicity, convolved with photometric errors and completeness as modeled by ASTs. The individual synthetic CMDs are linearly combined to form many possible SFHs. Each synthetic composite CMD is compared with the observed CMD via a Poisson maximum likelihood technique. The synthetic CMD that provides the best fit to the observed CMD is taken as the model SFH that best describes the data. For full implementation details, see \citet{Dolphin2002a}.

The fit quality is given by the \match\ \textit{fit} statistic: \textit{fit} = $-2\ln L$, where $L$ is the Poisson maximum likelihood. We estimate the $n\sigma$ confidence intervals as $n^2 \ge fit - fit_{\mathrm{min}}$; the 1$\sigma$ confidence interval includes all SFHs in a given region with $fit - fit_{\mathrm{min}} \le 1$, the 2$\sigma$ confidence interval includes all SFHs with $fit - fit_{\mathrm{min}} \le 4$, etc.

We use a fixed distance modulus of 24.47 \citep{McConnachie2005a}, a binary fraction of 0.35 with the mass of the secondary drawn from a uniform distribution, and a \citet{Kroupa2001a} IMF. We solve the SFH in 34 time bins covering a range in log time (in years) from 6.6 to 10.15 with a resolution of 0.1 dex except for the range of log(time) = 9.9 -- 10.15 which we combine into one bin. This time binning scheme was chosen to provide as much time resolution as possible while minimizing computing time. We found that using a finer time binning scheme with a resolution of 0.05 dex increased the computing time by at least a factor of two and only resulted in differences in the SFHs of \about1\%, which is much smaller than systematic and random uncertainties. We use the Padova \citep{Marigo2008a} isochrones with updated AGB tracks \citep{Girardi2010a}. The [M/H] range is [-2.3, 0.1] with a resolution of 0.1 dex. Because we are limited by the depth of the data, which does not reach the ancient main sequence turnoff, we also require that [M/H] only increases with time. We limit the oldest time bin to have [M/H] between -2.3 and -0.9 and the youngest time bin to have [M/H] between -0.4 and 0.1.

M31 contains significant amounts of dust \citep[e.g.,][]{Walterbos1987a, Draine2014a, Dalcanton2015a}, which, broadly speaking, can be described by three components: a mid-plane component due to extinction internal to M31 that dominates the older, well-mixed stellar populations, a foreground component due to Milky Way extinction, and a differential component that affects the star-forming regions. In addition to the SFH, \match\ allows two free parameters to describe the dust distribution: a foreground extinction (\av) and a differential extinction (\dav) which describes the spread in extinction values for the stars in each region. The differential extinction is a step function starting at \av\ with a width given by the value of \dav. While foreground extinction is expected to be relatively constant across the galaxy, differential extinction can vary significantly from region to region as they probe very different star-forming and stellar density environments. To determine the best fit to the data, we search extinction space to find the combination of \av\ and \dav\ that best fits the data. However, the distribution of dust is different for young stars and old stars \citep[e.g.,][]{Zaritsky1999a}. The step function differential extinction model provides a good fit to the main sequence (MS) component, but it cannot reproduce the post-MS stellar populations. 

\begin{figure*}[tb]
\centering
\includegraphics[width=\textwidth]{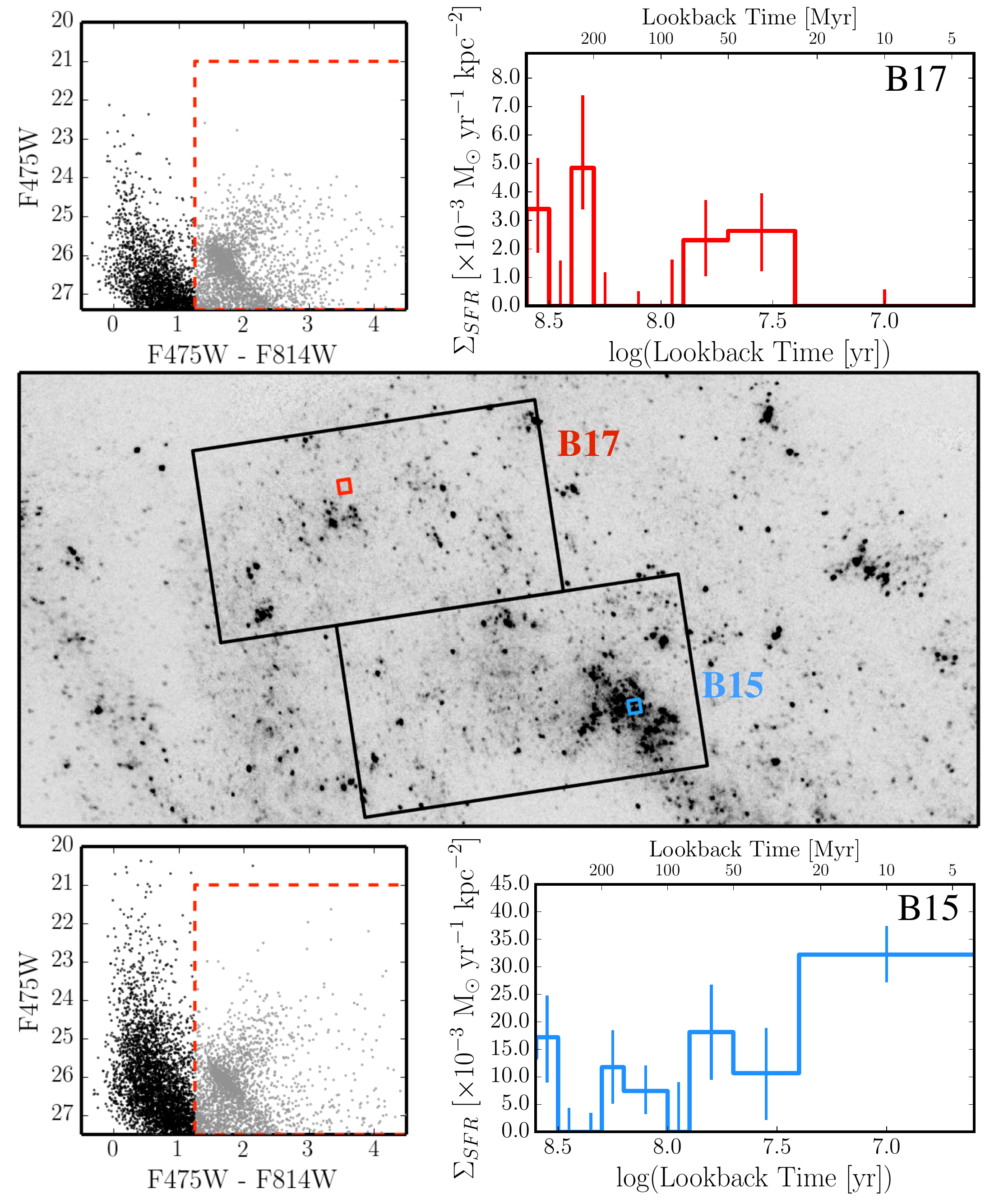}
\caption{Here we show example results for two different regions, one in Brick 15 sitting on a bright star-forming region (blue, lower) and the other in B17 in a more quiescent spot (red, upper). We have overlaid the outlines of Bricks 15 and 17 on top of a GALEX FUV image \citep{Gil-de-Paz2007a}. The individual region outlines are also shown. For each region, we show the CMD, with the stars that we fit in black and those that we don't in gray and outlined with red dashes, and the SFH to illustrate the differences on these small scales. The SFHs have been binned into 25 Myr increments below 100 Myr and 50 Myr increments above 100 Myr, as the native time resolution (0.1 dex) allows. Both of these regions have fairly recent SF as can be seen by the well-defined MS in each CMD, though the blue region (B15) shows ongoing and more intense SF. Surface densities are calculated using deprojected areas and assuming an inclination of 77$^\circ$ \citep{Brinks1984b}.}
\label{fig:image_sfh_cmd}
\end{figure*}

\begin{figure*}[t]
\centering
\includegraphics[width=\textwidth]{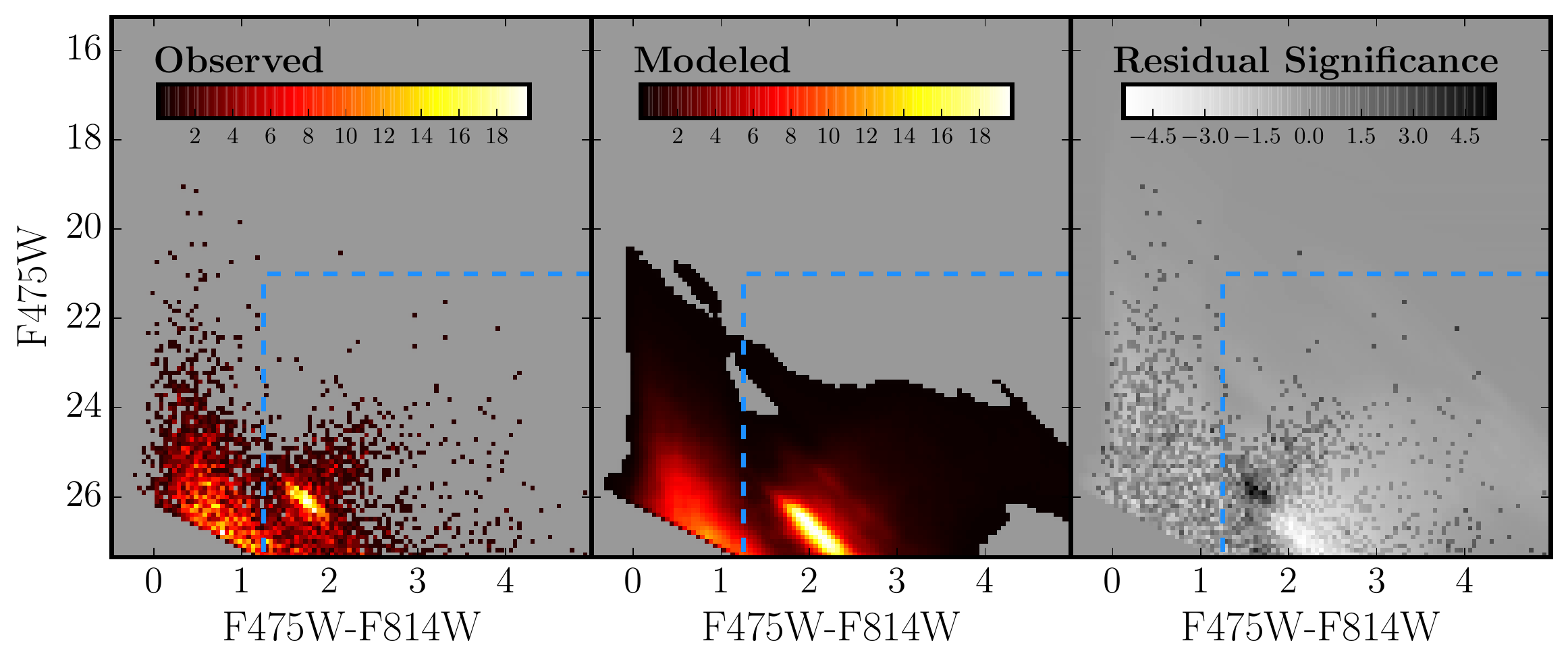}
\caption{The fit of a region in Brick 15. This is the same region as shown in Figure \ref{fig:image_sfh_cmd}. The left panel shows the data, the middle shows the best-fit model for the data, and the right shows the significance of the residuals (the observed CMD minus the modeled CMD, weighted by the variance) between the model and the data. All data inside the region outlined by the blue dashed line is masked out of the fit. In the left two panels, the color bars indicate the number of stars in each bin. In the right panel, the color bar indicates the significance of the residuals. In the residuals panel, darker colors generally indicate more stars in the data than in the model, lighter colors indicate more stars in the model than in the data. The lack of systematic residuals in the area that we fit is an indication that the model fits the data well.}
\label{fig:pg}
\end{figure*}

We mitigate the effects of dust on our SFHs by simplifying the fitting process such that we exclude the redder portions of the CMD from the fit. Specifically, we have adopted the cuts in \citet{Simones2014a}, excluding all stars with F475W-F814W$>$1.25 and F475W$>$21 (shaded regions of the CMDs in Figures \ref{fig:image_sfh_cmd} and \ref{fig:pg}). This prevents contamination from the older populations. We therefore avoid extinction-related complications by excluding the RGB and the red clump, which is often poorly fit with a single step function, and which is not relevant when calculating the recent SFH.  

We note that age-metallicity degeneracy is an important concern in any kind of SFH work. When modeling composite CMDs, it primarily affects the RGB \citep[e.g.,][]{Gallart2005a}, which we do not fit in this analysis. Instead, the vast majority of stars in the CMD are main sequence stars, for which the age-metallicity degeneracy is negligible compared to typical photometric uncertainties. In addition, the metallicity gradient of M31 has been extensively studied and found to be very shallow \cite[e.g.,][Gregersen et al. in prep]{Blair1982a, Zaritsky1994a, Galarza1999a, Trundle2002a, Kwitter2012a, Sanders2012a, Balick2013a, Lee2013a, Pilyugin2014a}, and the age range we are fitting is small. As a result, we are not concerned that the age-metallicity degeneracy  affects the results in this paper.

We compute the SFH in 450 regions per brick for 21 of the 23 bricks in the PHAT survey. To determine the best-fit SFH, we solve multiple SFHs with different combinations of \av\ and \dav, where the best-fit SFH is chosen to be the one whose combination of \av\ and \dav\ minimizes the fit value as given by the maximum likelihood technique. Consequently, for each of our regions, we must compute many possible SFHs. We minimize the total number of SFHs that must be run using an optimization scheme to limit the size of (\av, \dav) space that must be searched, as discussed in Appendix \ref{app:avdav}. Based on this optimization, we set a constraint that \av\ + \dav\ $\le$ 2.5. In each region, we run a grid of SFHs in (\av, \dav) space with a step size of 0.3 over the range of \av\ = [0.0, 1.0], also requiring that \av\ + \dav\ $\le$ 2.5. We take the resulting SFH with the best fit, determine the two-sigma range around that best (\av, \dav) pair, and then sample the grid in that region down to a finer spacing of 0.1 in \av\ and \dav. Not only does this ensure that we are finding the global minimum, but it also allows us to account for the uncertainty in extinction in the SFHs by including all fits in the result. In addition, the extinction parameters provide us with an additional method to verify our results, as we discuss in Section \ref{subsec:SFH_dust}.

As an example, in Figure \ref{fig:image_sfh_cmd} we plot the CMDs and SFHs for two of the regions found near the 10-kpc ring. We show the location of these regions over-plotted on a \textit{GALEX} FUV image \citep{Gil-de-Paz2007a}. The top region, in Brick 17, is located just off of a spur of the 10-kpc ring. The region itself shows very little FUV emission, and the resulting SFH is sparse with only moderate SFRs at all times. The lower region falls directly on an OB association \citep[OB 54;][]{vandenBergh1964a} in Brick 15. As expected, there is elevated, on-going SF in that region over at least the past 100 Myr. The CMDs for each of these regions show a well-defined MS. The consequences of dust are very evident in the region but are most easily seen in the part of the of the CMD that we do not fit, where the red clump is elongated along the reddening vector.

We note that the Padova isochrones do not include tracks younger than 4 Myr. In the resulting SFHs, we renormalize the SFR in the youngest time bin to reach the present day (0 Myr), conserving the total mass formed in that time bin. 

In Figure \ref{fig:pg}, we show the observed CMD, the best-fit modeled CMD, and the significance of the residuals (the observed CMD minus the modeled CMD with a weighting determined by the variance) for a region in Brick 15. We fit all stars that are outside the blue dashed region, which are primarily MS stars, with a smattering of short-lived blue helium-burning stars. The residuals show no distinct features, which means the model is a good fit.

\subsection{Extinction}
\label{subsec:SFH_dust}

In this section we discuss how we incorporate IR-based dust maps as a prior in determining our dust parameters. After determining the best-fit SFH in each region as described in Section \ref{subsec:SFH_fit}, we conducted additional verification by examining the map of total dust, \av\ + \dav. 

We found that there were a handful of regions in which the best-fit required large amounts of dust, at or very close to the limit of 2.5 mag in spite of there being no evidence for SF within the last 100 Myr, based on a lack of luminous MS stars and low SFR averaged over the most recent 100 Myr. In these regions, we would expect very little dust because there are no dust-enshrouded young stars. Bad fits result in these cases because there are few stars in the MCD fitting region that can be used to constrain the dust. 

To examine this discrepancy, we compared our dust parameters with the total dust mass in each region, as measured by \citet{Draine2014a}. We correct the low-SFR, high-dust regions by constructing a prior on \av\ + \dav\ based on the \citet{Draine2014a} dust mass maps and multiplying the prior by the likelihood calculated by \match. The details of the prior are described in Appendix \ref{app:prior}. After applying the prior, we compute new fits for all of the SFHs measured in each region. We use these new fits to go back through \av, \dav\ parameter space to be sure that we properly sampled 2-$\sigma$ space around the new best-fits for each region. As a result, we were able to constrain the \match\ dust parameters in the regions of very-low SFR that are not properly anchored in the CMD analysis. Ultimately, because application of the prior primarily affects the very-low SFR regions, this processing did not significantly affect the SFH results of this paper.

\subsection{Uncertainties}
\label{subsec:SFH_uncertainties}

There are three significant sources of uncertainties that affect the measured SFHs: random, systematic, and dust. In this section, we discuss each source of uncertainty in turn. 

First, we consider random uncertainties. The random uncertainties are dominated by the number of stars on the CMD and are consequently larger for more sparsely populated CMDs. Random uncertainties were calculated using a hybrid Monte Carlo process \citep{Duane1987a}, implemented as described in \citet{Dolphin2013a}.  The result of this Markov Chain Monte Carlo routine is a sample of 10,000 SFHs with density proportional to the probability density, i.e., the density of samples is highest near the maximum likelihood point.  Error bars are calculated by identifying the boundaries of the highest-density region containing 68\% of the samples, corresponding to the percentage of a normal distribution falling between the $\pm 1\sigma$ bounds. This procedure provides meaningful uncertainties for time bins in which the best-fitting result indicates little or no star formation.

Next, we consider the systematic uncertainties. Systematic uncertainties reflect deficiencies in the stellar models \citep[i.e., uncertainties due to convection, mass loss, rotation, etc.;][]{Conroy2013a} such that different groups model these parts of stellar evolution differently, which leads to discrepant results for the same data, depending on the stellar models used \citep{Gallart2005a, Aparicio2009a, Weisz2011a, Dolphin2012a}. These uncertainties primarily affect older populations that have evolved off the MS. The stellar models of the various groups generally agree quite well for MS stars that dominate our adopted fitting region.

Because we have used the same models across the whole survey, all regions experience similar systematic effects. We have estimated the size of the systematics for a number of regions, covering the range of stellar environments within M31. We computed the systematic uncertainties by running 50 Monte Carlo realizations on the best-fit SFHs as described in \citet{Dolphin2002a}. For each run, we shifted the model CMD in $\log(T_{\mathrm{eff}})$ and $M_{\mathrm{bol}}$ by an amount taken from a random draw from a Gaussian with sigma listed in Table \ref{tab:completeness}. These shifts are designed to mimic differences in isochrone libraries. We then measured the resulting SFH. The range that contained 68\% of the distributions from all 50 realizations is designated as the systematic uncertainties. The relative size of the systematics varies greatly from region to region but is generally less than half the size of the random uncertainties in individual regions and increases at larger lookback time. There is also more variation in the relative size of the uncertainties in the ring features than in the outermost regions where both the stellar density and the SFR are low.

Finally, the variable internal dust content introduces uncertainties. We select the best-fit SFH by choosing the (\av, \dav) pair that maximizes the likelihood. However, there are regions where the difference between the fit values of the two most likely SFHs is very small (i.e., both SFHs are almost equally likely). We also sample to a minimum spacing of only 0.1 in \av\ and \dav, and thus may have determined a slightly different best-fitting SFH than if we had sampled \av, \dav\ space more finely. To account for these variations, we calculate our uncertainties due to the dust distribution by combining all SFHs measured in a given region and determining the range that contains 68\% of the samples. In this combination, the SFHs are weighted by their fit values such that the best-fit gets full weight and the $n\sigma$ fits are weighted by $e^{-0.5 \, n^2}$ (e.g., SFHs with fit values that are $2\sigma$ from the best-fit value are weighted by $e^{-0.5 \times 2^2}$, or $e^{-2}$).

A possible additional source of uncertainty is due to the choice of binary fraction, which is a free parameter in this analysis. We tested the effect of different binary fractions in two of our regions and found that the final fit is not very sensitive to binary fraction. This is because the inclusion of binaries in the model results in a color separation on the CMD that is washed out by dust. The fits and resulting SFHs are consistent with the uncertainties when choosing a binary fraction anywhere between about 0.2 and 0.7. This insensitivity of the SFH to binary fraction is consistent with more extensive tests presented in \citet{Monelli2010a}. Uncertainties due to binary fraction will be much smaller than those due to dust.

We note that we do not include the model systematics in our reported uncertainties. While there may be absolute uncertainties in the global SFR due to model uncertainties, the relative region-to-region uncertainties are dominated by the random and dust components.

\subsection{Choice of Region Size}
\label{subsec:SFH_regionsize}

To generate the spatially-resolved SFH of M31, we divide each of the brick-wide catalogs into regions that are approximately 100 pc (projected; 25\arcsec) on a side, assuming a distance of 783 kpc to M31. There were a few different considerations for this size. 

Regions of this size are of scientific interest because they bridge the gap between existing knowledge of Galactic pc scale SF \citep[e.g.,][]{Bate2009a, Schruba2010a} and SF in more distant galaxies on kpc scales \citep[e.g.,][]{Leroy2008a}. The resolution is also fine enough to resolve features such as large HII regions and giant molecular clouds.

While a finer grid would also be scientifically interesting, there are a couple of difficulties to consider. The main problem is that smaller regions would have insufficiently populated CMDs, increasing the random uncertainties of the SFHs to unacceptable levels. With our adopted \about100 pc bin size, the number of stars within the CMD fitting region ranges from 110 to 3900. In 93\% of the regions, we fit more than 500 stars, and in 80\% we fit more than 1000 stars. Additionally, the SFHs are computationally expensive to run. For each region, \match\ must be run multiple times to determine the (\av, \dav) pair that provides the best fit to the data. Moving to 50 pc size regions would have resulted in four times as many regions, significantly increasing the time needed to derive the SFHs. Even at our 100 pc grid size, deriving the SFHs and uncertainties for the entire sample required more than 500,000 CPU hours using XSEDE resources \citep{Towns2014a}.

Our overall technique is similar to that of \citet{Simones2014a}, who measured the SFHs of UV-bright regions within Brick 15 of the PHAT survey. Their goal was to convert the SFHs into FUV fluxes and compare with the observed fluxes in each star-forming region. About half of their regions had fewer than 500 stars on the part of the CMD they were fitting but they still found reasonable agreement between the modeled and observed fluxes. This agreement indicates that the CMD fitting routine is robust, even with a modest number of stars in the part of the CMD occupied by young stars.

\subsection{Reliability of the SFHs as a Function of Lookback Time}
\label{subsec:SFH_reliability}

\begin{figure}[tb]
\centering
\includegraphics[width=\columnwidth]{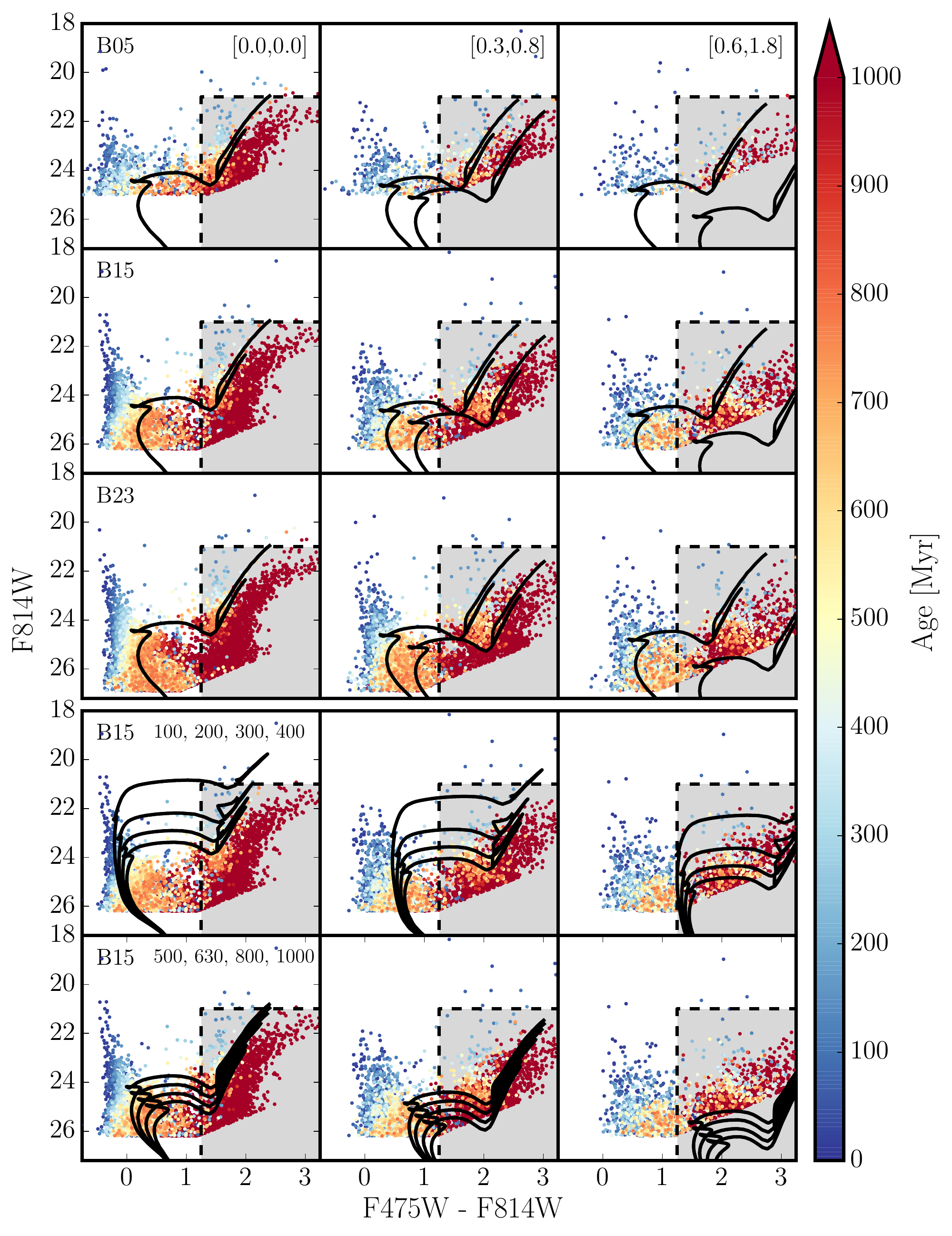}
\caption{Artificial CMDs over a range of extinction and photometric depth, assuming a constant SFH, solar metallicity, and three different extinction values; one low: [\av, \dav] = [0.0,0.0], one typical of our SFHs: [\av, \dav] = [0.3,0.8], and one high: [\av, \dav] = [0.6,1.8], shown in the left, middle, and right columns, respectively and labeled in the upper right corner of the top panel of plots. The brick number listed in the left panel of plots indicates the depth of the corresponding row. Within each panel, individual stars are color-coded by age. All stars older than 1 Gyr are dark red. In the top 3 rows, we have over-plotted 630 Myr, solar metallicity isochrones, one attenuated by \av, and the second attenuated by \av\ + \dav. Individual stars within the CMD will be reddened to somewhere between the two isochrones.  
Increased extinction reddens some stars off of the MS and into the region we do not fit in the SFH recovery process (shaded gray). The bottom two rows each have isochrones of 4 different ages over-plotted. The ages are listed in the left panel.}
\label{fig:cmd_age_iso}
\end{figure}

We  have excluded the red side of the CMD in our SFH recovery process, so consequently, our fits are not sensitive to old stellar populations. The exact age at which sensitivity is lost is set by the oldest stars observable on the MS, which varies with stellar density and dust extinction.

We perform two different tests to examine the sensitivity of our results as a function of time. First, we create artificial CMDs using \match. The CMDs are generated with a constant SFH and solar metallicity, while modeling observational uncertainties by using the results of the ASTs in each region. The results are shown in Figure \ref{fig:cmd_age_iso}, where we plot simulated CMDs. The youngest stars are shades of blue, and all stars older than 1 Gyr are red. Each of the top three rows shows a CMD at a different depth, where the top row is the shallowest CMD closest to the bulge (B05), and the third row is the deepest in B23. The brick numbers are indicated in the left panel. The columns display varying amounts of extinction, which is applied to the CMD in the same way we apply it to the model CMDs when recovering our SFHs. Extinction is labeled in the top panel by [\av, \dav]. The left column is un-reddened, the middle column shows the effect of the median extinction found in each of our regions ([\av, \dav] = [0.3, 0.8]), and the right column shows the upper limit of extinction allowed in our SFHs. We also check the sensitivity by over-plotting isochrones of a single age (630 Myr) and solar metallicity. The isochrone in the first column has not been reddened. In the second and third columns of the first three rows, we plot two isochrones, one extincted by \av\ and one extincted by \av\ + \dav. Individual stars can be extincted to anywhere between the two isochrones.

\begin{figure*}[h]
\centering
\includegraphics[scale=0.8]{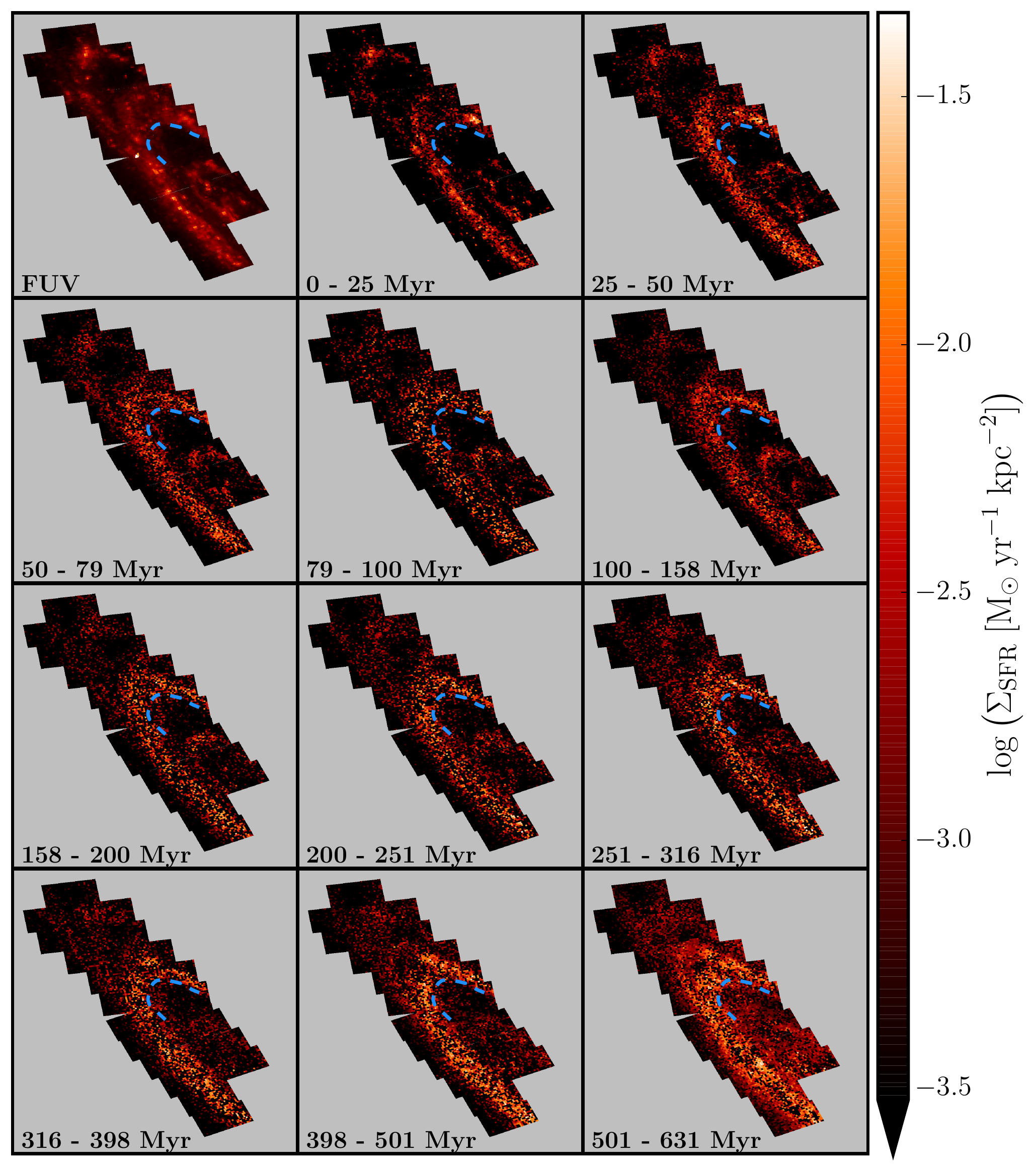}
\caption{Map of the SFH of M31 covered by the PHAT survey. Each region is colored according to its SFR surface density in each time bin. We have applied a lower cut at a \sigsfr\ of $10^{-4}$\sfrd\ in order to highlight structure in the image. The time range covered is shown in the lower left of each plot. The plot in the top left shows the \textit{GALEX} FUV image, smoothed to the same physical scale as our SFHs for better comparison. A blue dashed curve is over-plotted on each panel to aid the eye in recognizing structural change between timebins. The maps are oriented as in Figure \ref{fig:phat_area}. \textit{Note}: We have included the time bins from 400 Myr to 600 Myr for illustrative purposes only. As discussed in Section \ref{subsec:SFH_reliability}, the SFHs in these time bins are much less certain in the crowded inner regions and in the dusty ring than they are in the outer parts of the disk.} 
\label{fig:global_sfh_map}
\end{figure*}

The variation in depth is large across the survey. The inner regions are very crowded and as a result the photometric depth is quite shallow. In Brick 05, where there is high crowding and moderate levels of extinction, we cannot detect many stars that are older than \about500 Myr. As we move further away from the center of the galaxy, crowding and extinction generally decrease and we can probe stars that are 700 -- 1000 Myr old (row 3 of Figure \ref{fig:cmd_age_iso}). The exception, of course, is the 10-kpc ring, where extinction can be high; the CMDs in the second row of Figure \ref{fig:cmd_age_iso} mimic the conditions found in these regions. At low or mid-range extinction, we still see many stars on the MS that are at least 600 -- 700 Myr old. As we increase the extinction, many of these stars are reddened into the portion of the CMD that we do not fit, as can also be seen from the isochrones. Unreddened 630 Myr old stars are easily recovered; however, stars that have the highest level of extinction applied to them are entirely reddened into the neglected portion of the CMD. 

In the last two rows, the colored CMD is identical to that in the second row (B15 depth). Here, though, we plot solar metallicity isochrones of different ages: the fourth row has isochorones of 100, 200, 300, and 400 Myr while the bottom row shows isochrones of 500, 630, 800, and 1000 Myr. From these tests, we can see that there will be a number of regions that are reliable back to 1 Gyr, while others that are more extincted may only have a handful of stars that are more than \about500 Myr old. 

Ultimately, due to the wide variety of stellar environments found across the survey, not all bricks can reliably cover the same age range. The inner regions are much more crowded and have shallower completeness limits than the outer regions; as a result, the age limit for regions closest to the bulge is only 400 Myr. Regions that fall within the ring and spiral arm features are much more extincted than other regions; although the completeness limit is deeper, stars may be reddened into the region of the CMD that we do not fit. These regions have an age limit of no more than 500 Myr. In the outer regions where extinction and crowding are less, the completeness limit is much deeper and the SFHs are reliable back to 700 -- 1000 Myr. We therefore choose a conservative age limit of 400 Myr for consistency across the survey. For all scientific analysis, we examine only the most recent 400 Myr. In some of the plots that follow, we include results back to 630 Myr for illustrative purposes, with the caveat that they are only relevant for the outermost regions of the survey.

\section{Results}
\label{sec:results}

We now present the results of the SFH fitting process.

\subsection{Star Formation Rate Maps}
\label{subsec:sfr}
We combine the single best-fit SFHs in each region into maps of \sigsfr\ as a function of time. As an ensemble, they reveal the recent SFH of M31 in the PHAT footprint over the last 630 Myr. This SFH is shown in Figure \ref{fig:global_sfh_map}. Each panel displays the SFR surface density within the time range specified in the lower left corner. We note that these time bins are not the native resolution of $\Delta(\log t)=0.1$; instead, we have binned the results within the most recent 100 Myr into \about25 Myr bins. This helps to expose the continuous structure, especially in the most recent 25 Myr when SF is very low across the galaxy. The regions are colored according to $\Sigma_{\rm SFR}$ in that region during the given time range. To more clearly illuminate the structure, we have placed a lower limit on the SFRs visible within this map such that all regions with $\Sigma_{\rm SFR}$ lower than $10^{-4}$\sfrd\ are colored black. We have also over-plotted a blue dashed line on each image to aid the eye in recognizing structural evolution between time bins. There is little large-scale change over the last \about500 Myr. The upper left panel shows the \textit{GALEX} FUV image smoothed to the same physical scale as our SFHs, which shows excellent morphological agreement with the most recent time bins. This agreement strengthens confidence in our SFH maps given that measurements made in \about9000 completely independent regions reproduce coherent large-scale structure seen in a well-established SFR tracer. The relationship between the SFR and the FUV flux will be examined in an upcoming paper (Simones et al. in prep.).

\begin{figure}[]
\centering
\includegraphics[width=\columnwidth]{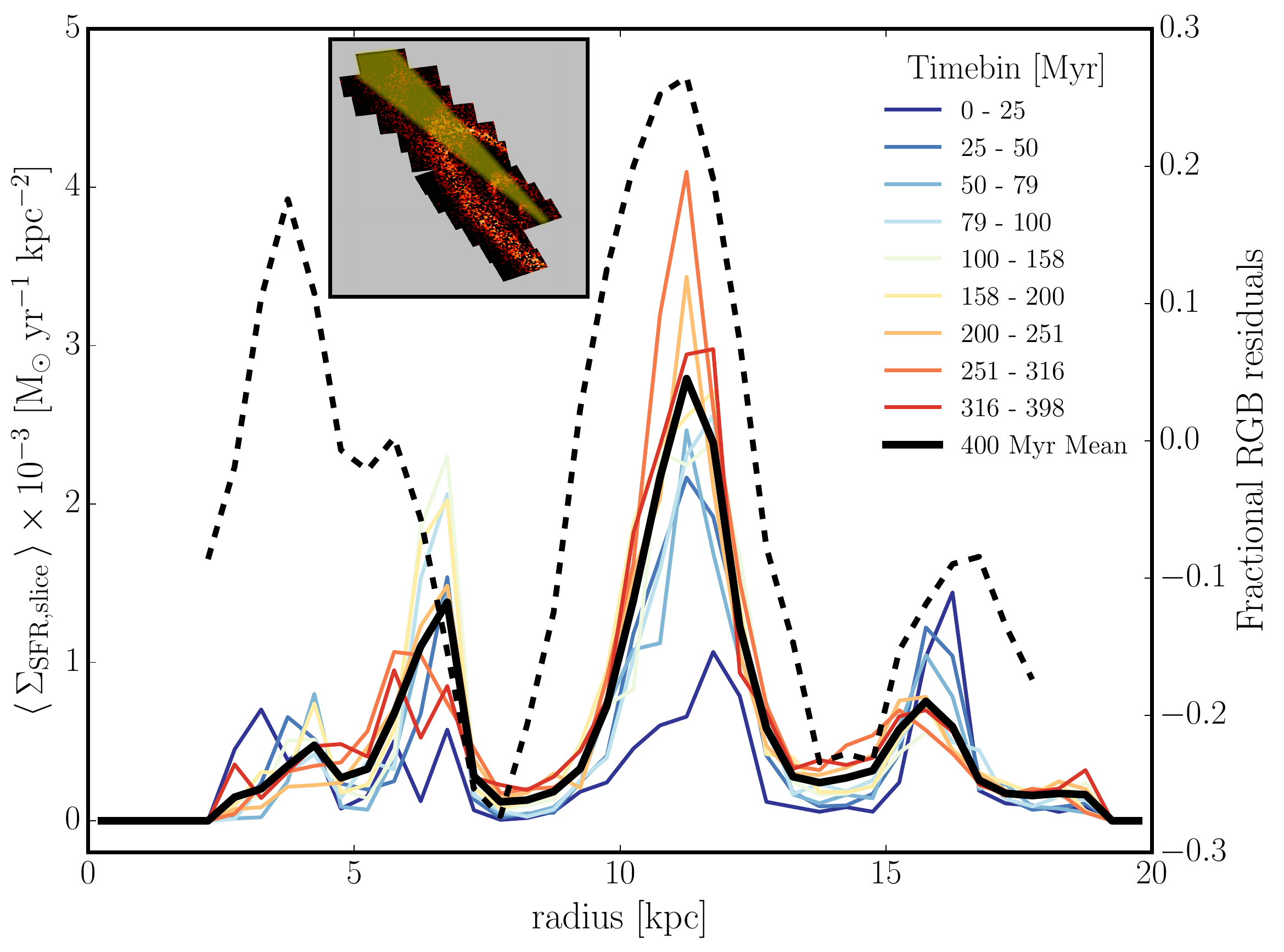}
\caption{SFR surface density as a function of radius and time. Here we plot the SFR surface density in all regions within a 40 degree arc about the major axis (shaded yellow in the inset) per time bin. We have binned the radius in 0.5 kpc bins. The SFR surface density in each radius bin is the mean of all regions that fall within that bin. The thick black line shows the overall mean. The dashed line shows the fractional RGB residuals (Seth et al., in prep.) with values given on the right axis.}
\label{fig:sigma_radius}
\end{figure}

The SFHs reveal large-scale, long-lasting, coherent structures in the M31 disk. There are three star-forming ring-like features; a modest inner ring at \about5 kpc, the well-known 10-kpc ring, and an outer, low-intensity ring at \about15 kpc that partially merges with the 10-kpc ring due to a combination of projection effects and a possible warp that is visible in HI \citep[e.g.,][]{Brinks1984a, Chemin2009a, Corbelli2010a}. These rings have been observed previously in \textit{Spitzer}/Infrared Array Camera (IRAC) images \citep{Barmby2006a} and \textit{Spitzer}/Multiband Imaging Photometer (MIPS) images \citep{Gordon2006a}, as well as in atomic \citep{Brinks1984a} and molecular gas images \citep{Nieten2006a}. There is also an observed over-density of red giant branch (RGB) stars in the 10-kpc ring \citep{Dalcanton2012a}. Recovery of these coherent, well-known features is further confirmation that our method is robust.

One of the most remarkable features of this SFH is that the 10-kpc ring is visible and actively forming stars throughout the past \about500 Myr. Although SF occurs at a low level in the outer regions at all times, SF in the ring feature at 15 kpc is most concentrated starting at 80 Myr ago. This is likely due to SF in OB 102, which has had an elevated SFR compared to its surroundings over the last \about100 Myr \citep{Williams2003b}. The inner ring feature at $\sim$5 kpc is also visible, and though there appears to be SF in that ring distinct from the surrounding populations, that feature gains definition 200 Myr ago but has largely dispersed in the last 25 Myr.

\begin{figure*}[tb]
\centering
\includegraphics[scale=0.8]{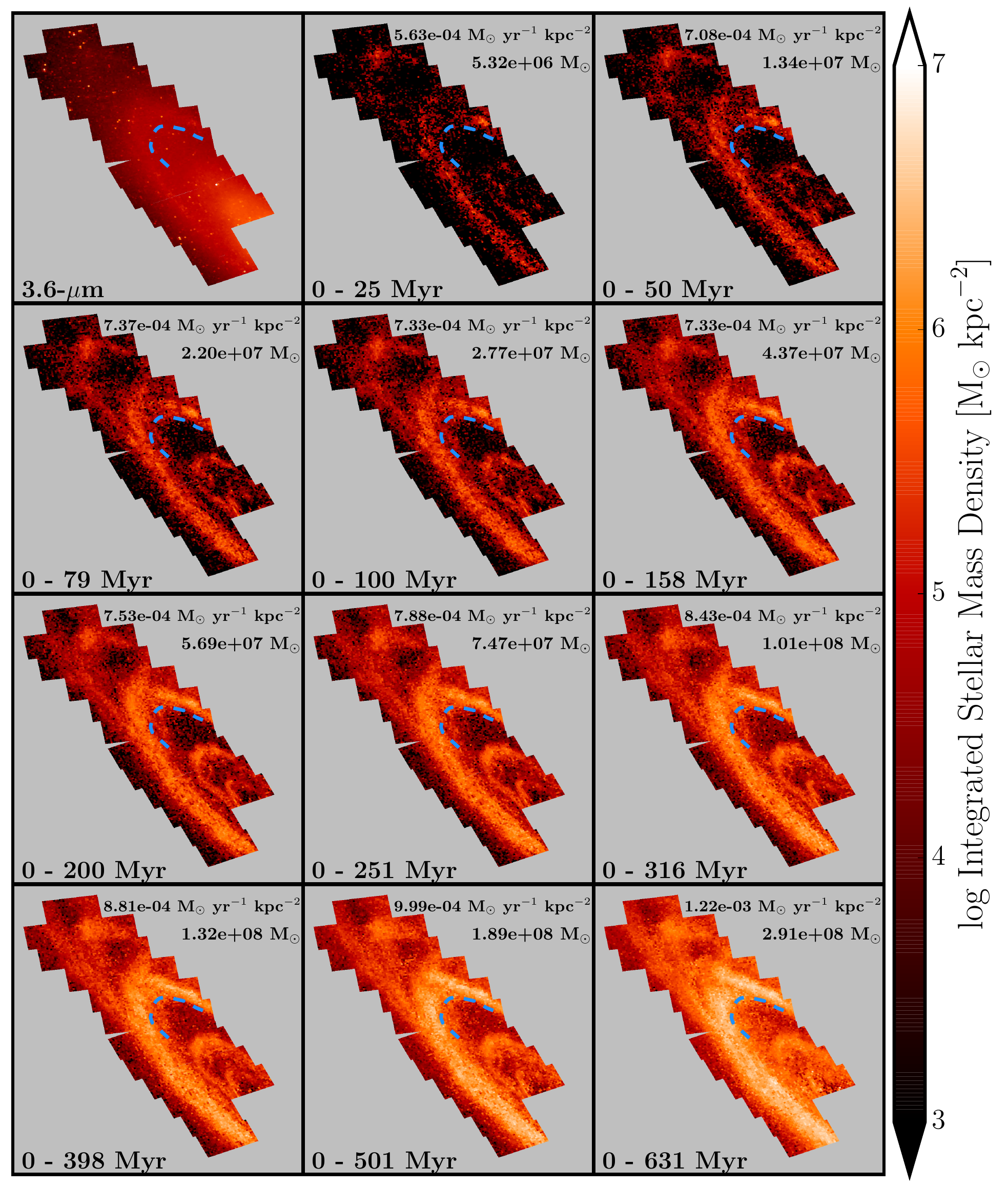}
\caption{Map of the cumulative mass formed. Each region is colored according to the mass formed in that region, integrated from the present day, and scaled by the area of the region. The time range covered is shown in the lower left of each plot. In the upper right of each plot, we indicate the SFR surface density and the total mass formed over that time range. We have chosen to present the maps in this way in order to better highlight the structure of the most recent mass formed. Most of the mass in M31 was in place by at least 1 Gyr ago. We scale the maps to the mass formed in the last 630 Myr in order to highlight structure. The maps are oriented as in Figure \ref{fig:phat_area}.  \textit{Note}: We have included the time bins from 400 Myr to 600 Myr for illustrative purposes only. As discussed in Section \ref{subsec:SFH_reliability}, the SFHs in these time bins are much less certain in the crowded inner regions and in the dusty ring than they are in the outer parts of the disk.}
\label{fig:global_mass_map}
\end{figure*}

We further investigate these trends in Figure \ref{fig:sigma_radius} where we plot the average SFR surface density as a function of radius in each time bin for a subset of the regions along the major axis. We determine the distance to the center of each region and then divide the regions into bins of 0.5 kpc. The three rings at \about5 kpc, \about10 kpc, and \about15 kpc are clearly visible as peaks in \sigsfr, providing further indication of ongoing (if low) SF in the ring features over 500 Myr. This trend is further supported by evidence of RGB star residuals in the ring features \citep[Seth et al., in prep][]{}, as shown by the thick dashed line. There is an over density of RGB stars in the 10-kpc ring and in the inner ring where our oldest time bin shows the highest SF. The plot reveals that not only is the 10-kpc ring long-lived, it has also remained mostly stationary in galactocentric radius over 500 Myr, a result that has implications for the origin of the ring, which we discuss further in Section \ref{subsec:ring}.

\subsection{Mass Maps}
\label{subsec:mass}

In Figure \ref{fig:global_mass_map}, we show the evolution of recent mass growth in the galaxy from 630 Myr to the present. The upper left panel shows the 3.6 \micron\ image, which is a rough estimation of the total mass of a galaxy, smoothed to the same spatial resolution as our SFHs. The next 11 panels show the total mass formed over a given time range (the same as those in Figure \ref{fig:global_sfh_map}). The upper middle panel shows the mass that formed in the last 25 Myr while the bottom right panel shows all of the mass formed in the last 630 Myr. The time range is written in the lower left of each plot. In the upper right, we have indicated the total mass formed and the average SFR over that time range. 

Most of the SF in M31 occurred much earlier than the timescale probed by these SFHs; consequently, the amount of mass accumulated over the last \about500 Myr has been minimal. As can be seen in the upper left image in Figure \ref{fig:global_mass_map}, the older stars are distributed fairly uniformly. This means that the structure that we see in our integrated mass maps only appears within the last \about2 Gyr (more recent than the ages of stars that dominate the emission at 3.6 \micron). Over the time range of our SFHs, SF has been confined primarily to the rings. 

\begin{figure}[tb]
\centering
\includegraphics[width=\columnwidth]{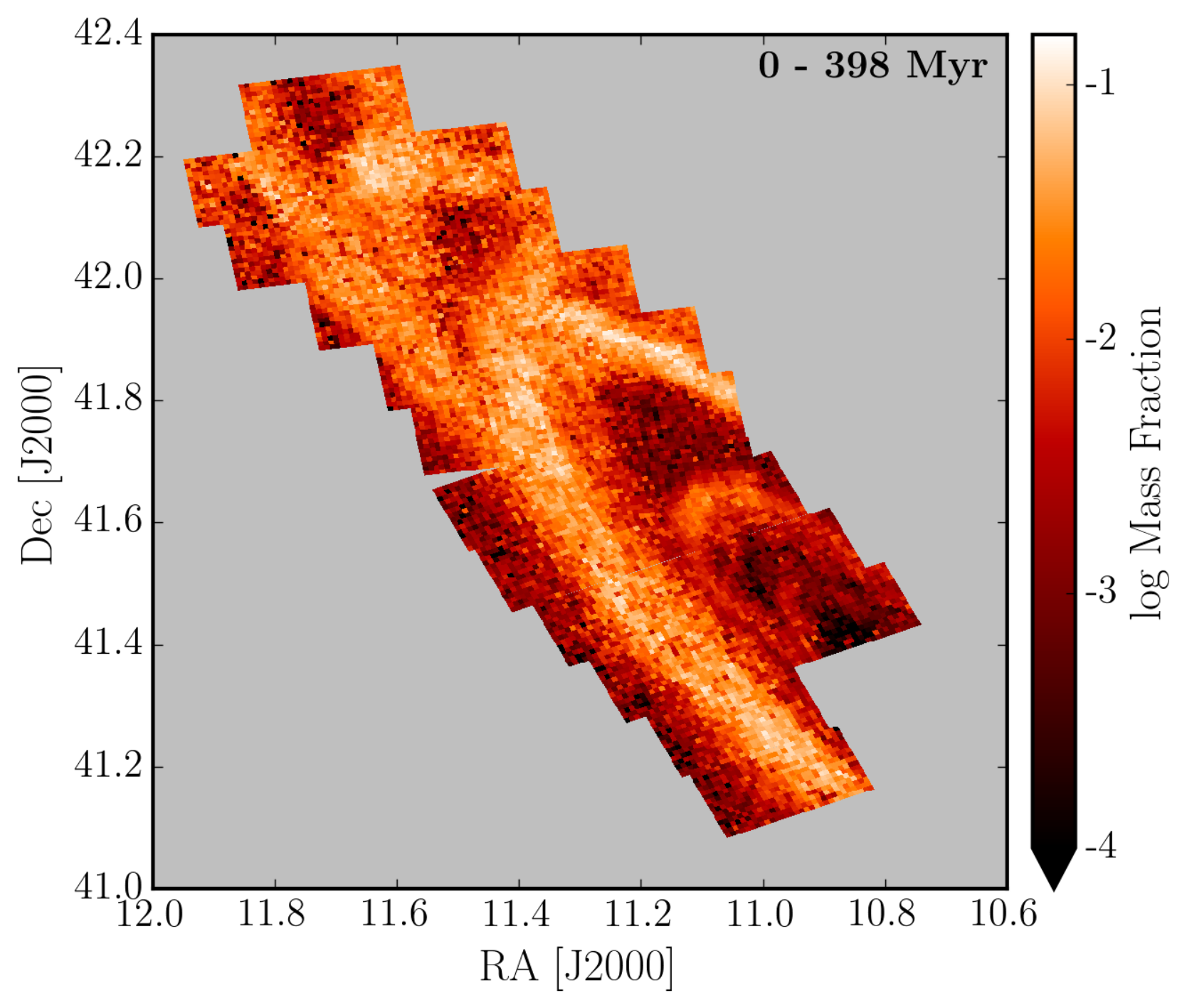}
\caption{Map of the fraction of mass formed in the last 400 Myr compared to the total mass as inferred from the 3.6 \micron\ image \citep{Barmby2006a}. The amount of mass formed over the time range covered by these SFHs is very small, as expected, with the highest fractions coming in the 10 and 15-kpc ring features. The map is oriented as in Figure \ref{fig:phat_area}.}
\vspace{1mm}
\label{fig:mass_fraction_map}
\end{figure}

We further examine the impact of SF in the last \about500 Myr by looking at the fraction of mass formed during this time. In Figure \ref{fig:mass_fraction_map}, we plot the fraction of mass formed in the last 400 Myr to the total mass in each region. We derive the total stellar mass by multiplying the 3.6 \micron\ image \citep{Barmby2006a} by a constant mass-to-light ratio of 0.6 \citep{Meidt2014a}.  Only \about0.8\% of the total stellar mass of this section of the galaxy formed in the last 400 Myr, which is \about3\% of a Hubble time. On a region-by-region basis, the mass fraction reaches a maximum of \about7\% and is highest in the ring features, specifically in the two outer features, where most of the gas is located (see Figure \ref{fig:sfr_fuv_hi}). Consequently, though the SFR in the ring in the last \about500 Myr is much higher than that in the rest of the galaxy, it must still be very small relative to past SFRs.

\section{Discussion}
\label{sec:discussion}

\begin{figure*}[]
\centering
\includegraphics[width=\textwidth]{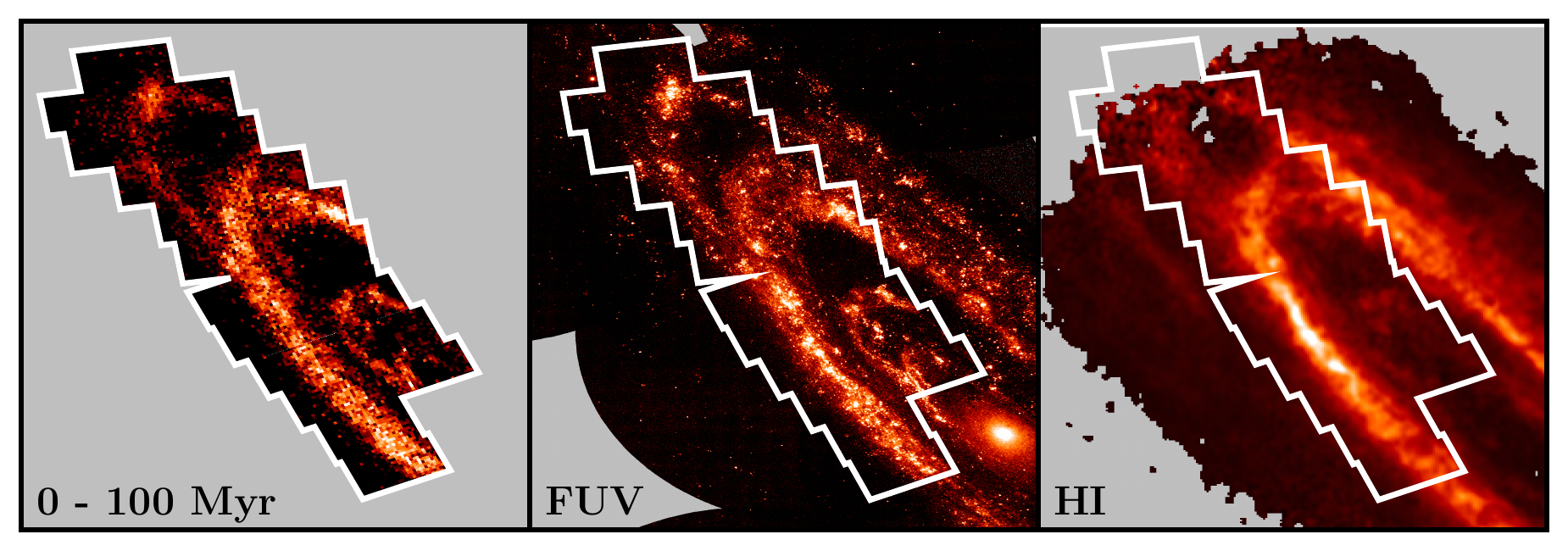}
\caption{Maps of the average SFR in the last 100 Myr as derived in this paper, \textit{GALEX} FUV \citep[][4.3\arcsec\ resolution]{Gil-de-Paz2007a}, and HI \citep[][45\arcsec\ resolution]{Brinks1984a}. There is good morphological agreement between the SFR map and the FUV map, which is sensitive to the same time scale. It is also clear that the regions of highest SF, primarily in the 10-kpc ring, coincide with the regions of highest HI content. Orientation of the images is as in Figure \ref{fig:phat_area}.}
\label{fig:sfr_fuv_hi}
\end{figure*}

We have presented maps of SF and mass evolution in M31 that show rich structure with ongoing SF and evolution. Much of this structure has been observed in maps of M31 in other tracers. As an example, in Figure \ref{fig:sfr_fuv_hi}, we plot our SFR map averaged over the last 100 Myr, next to maps of FUV flux \citep{Gil-de-Paz2007a} and HI \citep{Brinks1984a}. All three maps show M31's ring structures. In addition, we note the good agreement between the 100 Myr SFR map and the map of FUV flux, which is sensitive to SF within the last 100 Myr. These regions of high SFR and flux are also broadly consistent with areas of largest gas content. We now discuss some of the features observed in these maps.

\subsection{The Recent PHAT SFH}
\label{subsec:totalSFH}

The SFHs of the individual regions are extremely diverse. Some of the regions, such as those in the 10-kpc ring have SF that is ongoing and seemingly long-lived; others, such as those that fall in the outer parts of the galaxy, have generally quiescent SFHs in the past 500 Myr. At the native time resolution of 0.1 dex, the uncertainties on the individual SFHs in each region are significant. By combining all the SFHs, we can reduce the amplitude of the uncertainties in those time bins and obtain a more significant and constrained result for the total SFH. 

\begin{figure}[b]
\centering
\includegraphics[width=\columnwidth]{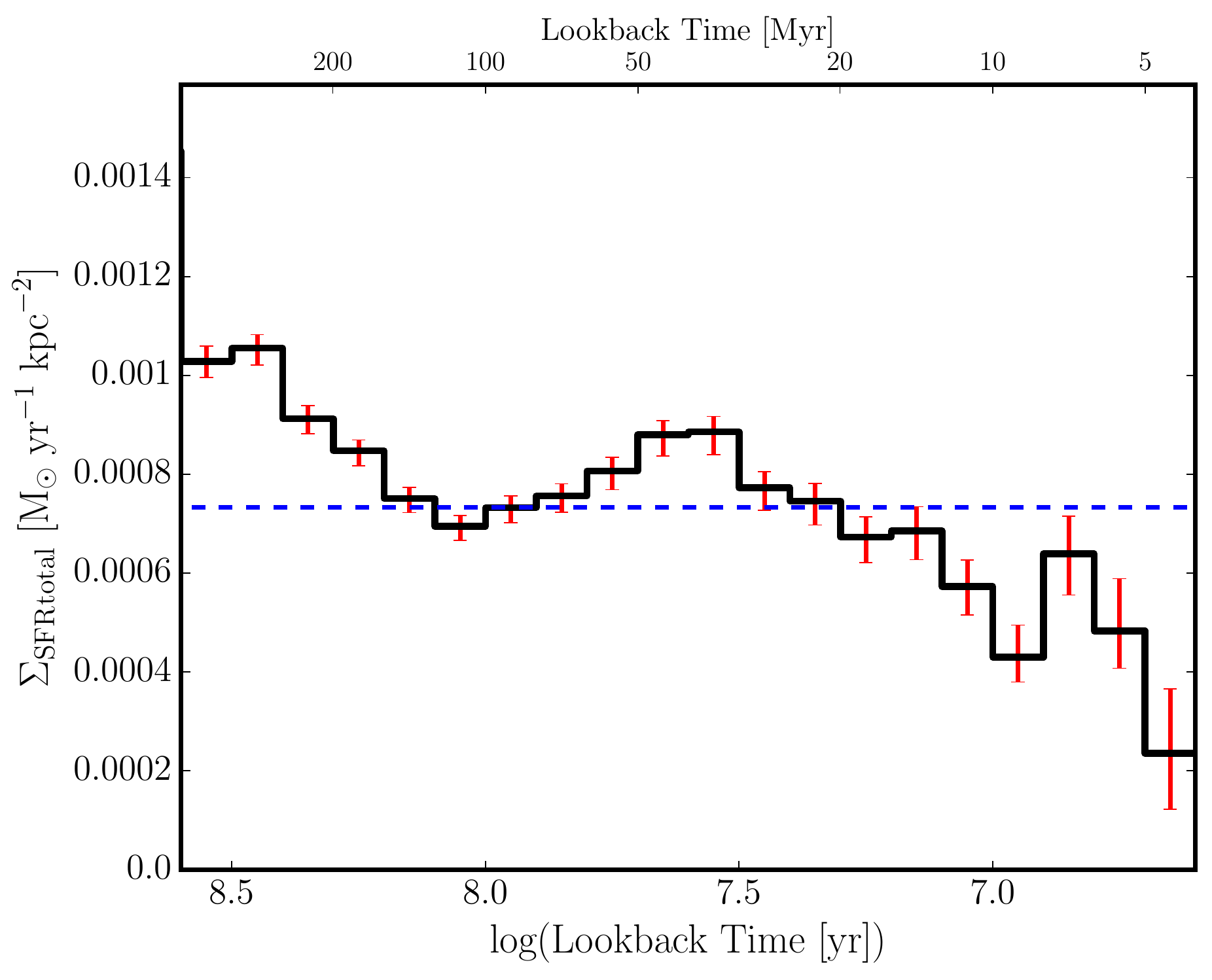}
\caption{The total SFH of M31 within the PHAT footprint over the last 400 Myr, combining the individual SFHs of each field. The dashed line (blue) shows the average SFR density over the most recent 100 Myr. The red error bars are a combination of the random uncertainties and the uncertainties in \av\ and \dav. The time bins are $\Delta(\log t)=0.1$, so the youngest bins cover considerably less linear time than do the older bins.}
\label{fig:global_sfh}
\end{figure}

\begin{figure*}[tb]
\centering
\includegraphics[width=\textwidth]{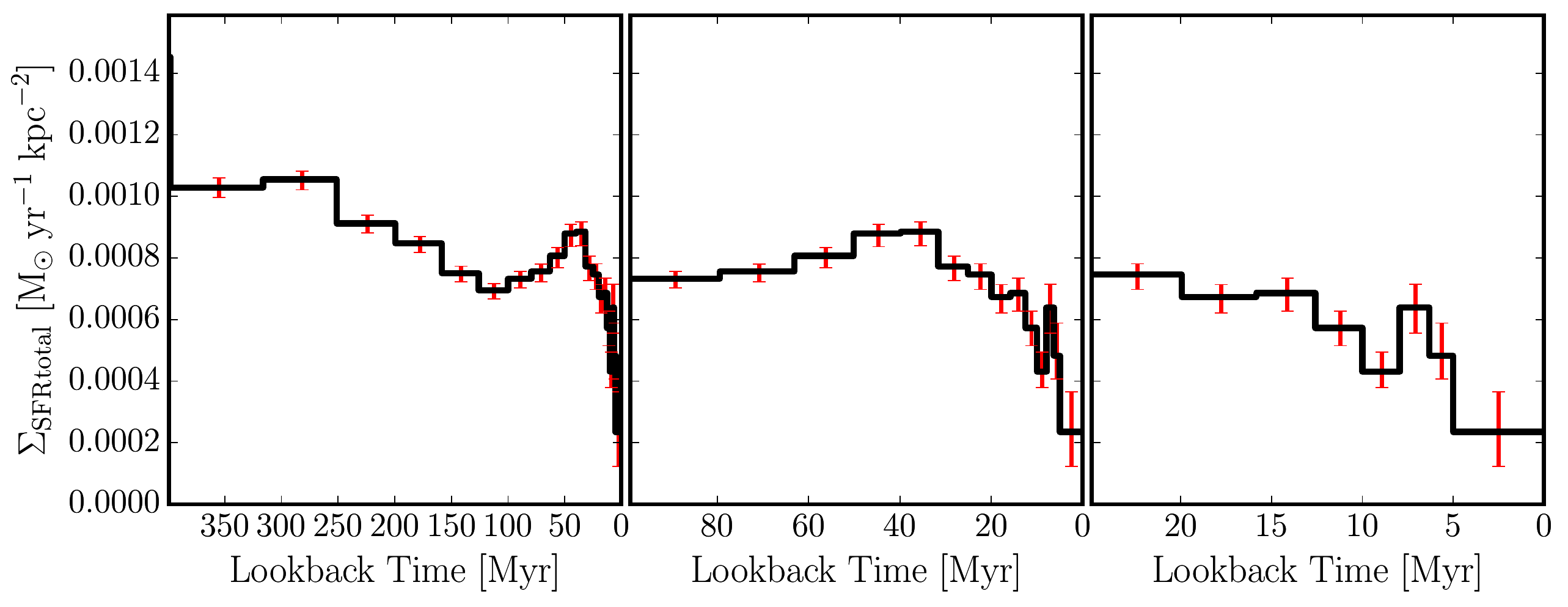}
\caption{Same as in Figure \ref{fig:global_sfh} but showing time bins on a linear scale. The first panel shows the SFH back to 400 Myr. The second shows the SFH over the last 100 Myr and the third panel shows the SFH within the last 25 Myr.}
\label{fig:global_sfh_linear}
\end{figure*}

We derive the total SFH within the PHAT footprint by integrating over all regions in Figure \ref{fig:global_sfh_map}. In Figure \ref{fig:global_sfh} we show the total SFR per time bin over the survey area. Figure \ref{fig:global_sfh_linear} shows the same SFH but with linear time bins over 3 different time ranges. The dashed blue line indicates the average SFR over the past 100 Myr. The uncertainties in each of these figures include the random component as well as uncertainties in the \av, \dav\ combination of the best-fit SFH of each region. There are also systematic uncertainties due to isochrone mismatch (see Section \ref{subsec:SFH_uncertainties}), which we do not include but are \about30\% in all time bins. The SFRs and uncertainties in each time bin are listed in Table \ref{tab:totalSFH}.

\begin{figure*}[b]
\centering
\includegraphics[width=\textwidth]{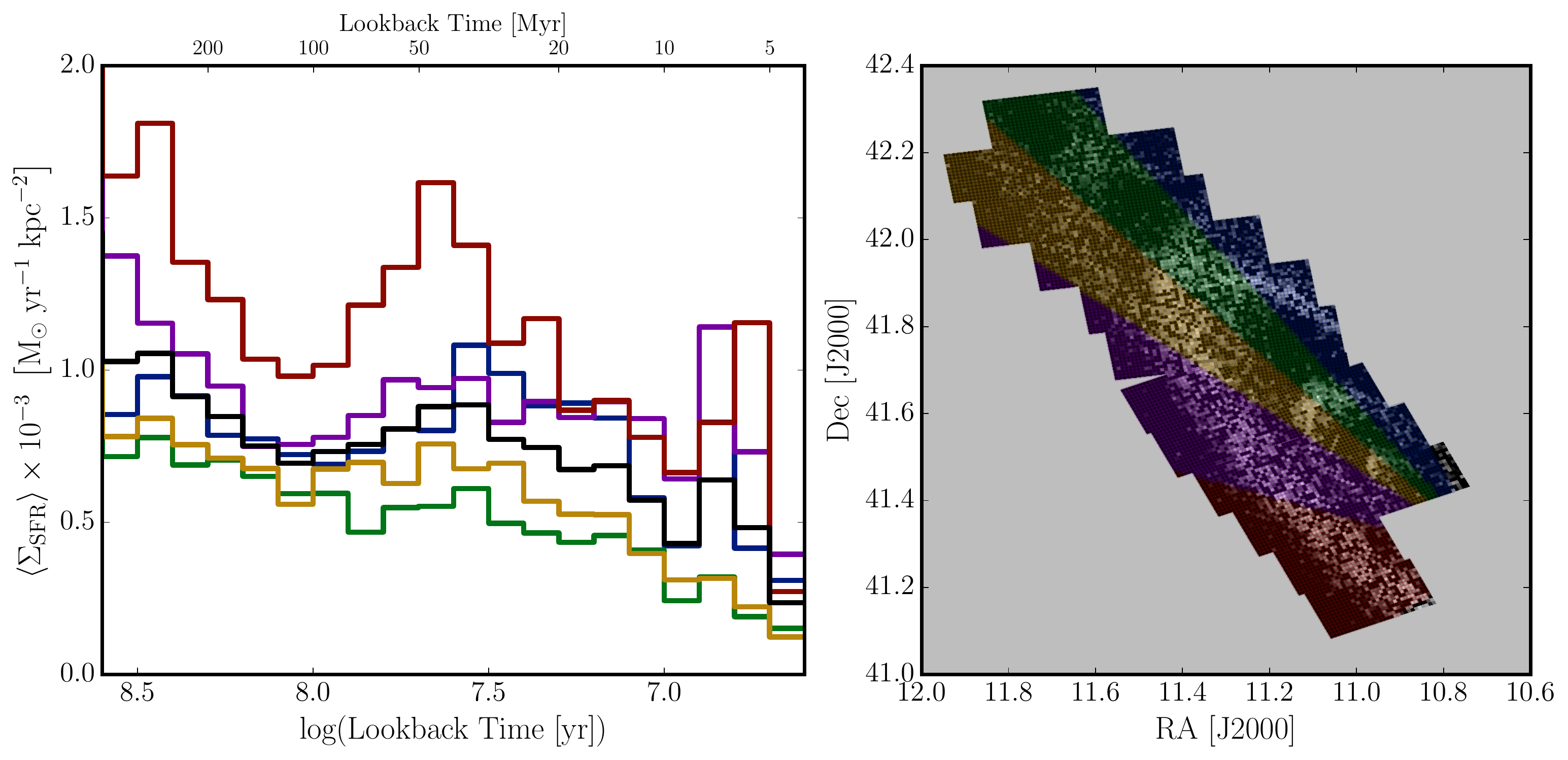}
\caption{SFH in slices about the major axis. On the left we show the average SFR per area in each time bin for all of M31 divided into 5 angular slices. The lines are colored according to angular slice. The same coloring is used in the right panel to indicate the angular slice that is represented. All slices cover 30 degrees, though they do not all contain the same number of regions. The black line shows the average SFR per time bin for the entire survey area.}
\label{fig:theta_sfh}
\end{figure*}

\begin{deluxetable}{ccc}
\tabletypesize{\footnotesize}
\tablecaption{Total PHAT SFH} 
\tablecolumns{3} 
\tablewidth{0pt}
\tablehead{ 
 \colhead{log(ti)} & 
 \colhead{log(tf)} & 
 \colhead{SFR}  \\ 
 \colhead{log(yr)} & 
 \colhead{log(yr)} & 
 \colhead{\sfr}
} 
\startdata
6.60 & 6.70 & 0.09$^{+0.04}_{-0.03}$ \\
6.70 & 6.80 & 0.18$^{+0.03}_{-0.02}$ \\
6.80 & 6.90 & 0.24$^{+0.02}_{-0.02}$ \\
6.90 & 7.00 & 0.16$^{+0.02}_{-0.01}$ \\
7.00 & 7.10 & 0.22$^{+0.02}_{-0.02}$ \\
7.10 & 7.20 & 0.26$^{+0.01}_{-0.02}$ \\
7.20 & 7.30 & 0.25$^{+0.01}_{-0.01}$ \\
7.30 & 7.40 & 0.28$^{+0.01}_{-0.01}$ \\
7.40 & 7.50 & 0.29$^{+0.01}_{-0.01}$ \\
7.50 & 7.60 & 0.34$^{+0.01}_{-0.01}$ \\
7.60 & 7.70 & 0.33$^{+0.01}_{-0.01}$ \\
7.70 & 7.80 & 0.30$^{+0.01}_{-0.01}$ \\
7.80 & 7.90 & 0.29$^{+0.01}_{-0.01}$ \\
7.90 & 8.00 & 0.28$^{+0.01}_{-0.01}$ \\
8.00 & 8.10 & 0.26$^{+0.01}_{-0.01}$ \\
8.10 & 8.20 & 0.28$^{+0.01}_{-0.01}$ \\
8.20 & 8.30 & 0.32$^{+0.01}_{-0.01}$ \\
8.30 & 8.40 & 0.34$^{+0.01}_{-0.01}$ \\
8.40 & 8.50 & 0.40$^{+0.01}_{-0.01}$ \\
8.50 & 8.60  & 0.39$^{+0.01}_{-0.01}$ 
\enddata
\label{tab:totalSFH}
\tablecomments{The total SFH summed over all regions. The first column lists the start of each timebin and the second column lists the end of each time bin. The third column shows the total SFR in each time bin, integrated over the entire survey. To convert to SFR surface density ($\Sigma_{\rm SFR}$), divide the SFR by the total area of the survey, 378 kpc$^2$. The uncertainties represent the smallest range that contains 68\% of the probability distribution calculated fom the random and dust uncertainties. The first time bin extends to the present day.}
\end{deluxetable}

We look more closely at the total SFH to determine which regions contribute in each time bin. While the SFR has been generally declining over the last 600 Myr, we see a bump at \about50 Myr. Further examination reveals that this peak is not a galaxy-wide feature. To show this, in Figure \ref{fig:theta_sfh}, we plot the SFH in 5 angular slices in theta about the major axis. We see that the slice that falls down the major axis (green line) shows a declining SFR over all periods of time, i.e., there is no 50 Myr bump. The west-most slice (blue line) shows a slight peak at 50 Myr and if we move to the slice that is out along the minor axis (purple line), the SFR shows the most prominent 50 Myr bump. These slices contain bright OB associations and are dominated by ring regions. Consequently, it is likely that the 50 Myr peak in the total SFH is due to specific star-forming regions in the rings.

In Figure \ref{fig:global_sfh}, we have over-plotted a dashed line showing the average SFR over the past 100 Myr. The average SFR in the PHAT area over the past 100 Myr is 0.28$\pm$0.03\sfr, where the uncertainty is calculated from the tests of systematic uncertainties. To determine the average SFR over the past 100 Myr over the entire disk, we adopt a simple scaling argument. We use both the FUV and 24-\micron\ maps of M31 and select the D25 ellipse \citep{Gil-de-Paz2007a} as an aperture. We then measure the total flux within the entire aperture and the flux inside the PHAT footprint (without Bricks 1 or 3). In both cases, the fraction of the total flux inside the PHAT footprint is \about40\%. Therefore, to scale the PHAT SFR to the entire D25 aperture, we must multiply by a factor of 1/0.4 = 2.5.  We multiply the SFR by this scaling factor to yield a total SFR of \about0.7\sfr. Previous studies have examined the global SFR using resolved stars \citep[\about1\sfr;][]{Williams2003a}, 8 \micron\ \citep[0.4\sfr;][]{Barmby2006a}, FUV \citep[0.6--0.7\sfr;][]{Kang2009a}, H$\alpha$ (\about0.3\sfr;\citealt[][]{Tabatabaei2010a}, 0.44\sfr; \citealt[][]{Azimlu2011a}), and FUV + 24 \micron\ \citep[0.25\sfr;][]{Ford2013a}. Our result falls within this range, and is most consistent with methods that effectively average over longer timescales.

Although this simple scaling argument neglects possible variations in the SFR from one side of the disk to the other, we do not expect these variations to be large enough to dramatically affect the overall SFR. A complete analysis of the SFR measured via different tracers is beyond the scope of this paper but will be examined more closely in an upcoming paper (Lewis et al., in prep).

\subsection{Birthrate Parameter}

We further examine the total SFH by computing the birthrate parameter, $b$, which is defined as the ratio of the current SFR to the past averaged SFR:

\begin{equation}
	b = \dfrac{\mathrm{SFR}}{\langle \mathrm{SFR} \rangle}
\end{equation}

We calculated the past-averaged SFR by converting the 3.6 \micron\ image to mass, as described in Section \ref{subsec:mass} and setting the lifetime of the disk to the Hubble time.

\begin{figure*}[tb]
\centering
\includegraphics[width=\textwidth]{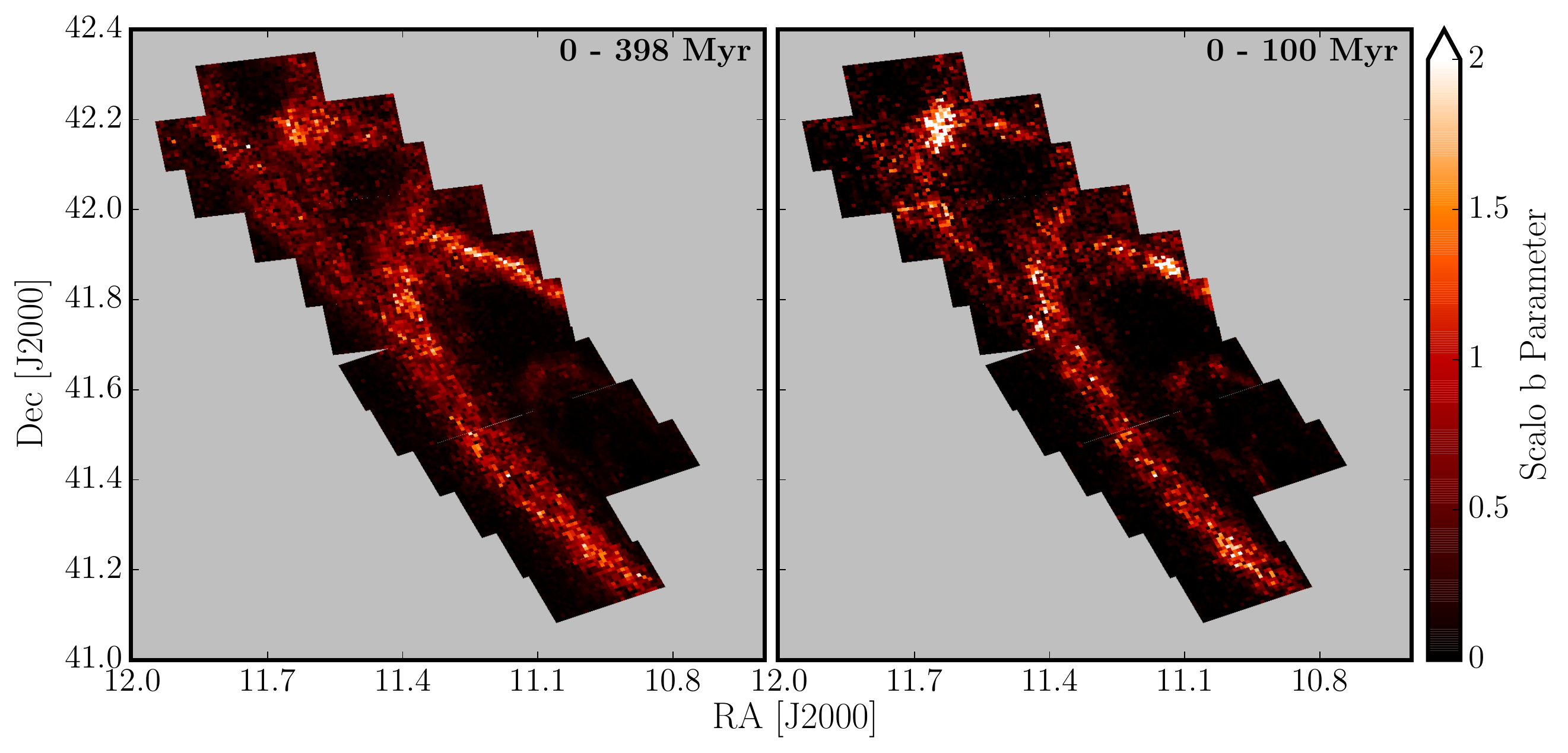}
\caption{Maps of the birthrate parameter $b$ over 400 Myr and 100 Myr timescales. The regions are colored according to the value of the birthrate parameter in that region and an upper limit has been set at $b=2$. The maps are oriented as in Figure \ref{fig:phat_area}. Both maps show $b>1$ in the 10 and 15-kpc rings. The OB associations in Bricks 15 and 21 are clearly visible in the right image as regions of elevated $b$.}
\label{fig:birthrate}
\end{figure*}

While the birthrate parameter is generally used to classify a galaxy as a whole as in either a star-bursting or quiescent phase, in this case we can examine $b$ for our individual regions. This gives us the ability to tell whether regions of current high SF, such as the 10-kpc ring, are currently undergoing unusually high SF relative to the rest of the galaxy (in which case $b$ should be large) or whether those regions have always had higher SFRs (in which case $b \le 1$). In Figure \ref{fig:birthrate}, we show a map of the birthrate parameter over two timescales: the lifetime of our SFHs (\about400 Myr) and 100 Myr.  The regions are colored according to $b$ on a linear scale. We see over both time periods that the rings at 10 and 15 kpc have higher values of $b$ than the rest of the galaxy. While $b$ is greater than unity over most of these ring features, there are very few regions where it rises above 2, which indicates that while SF is certainly elevated in those regions compared to the overall SFR, it would not be considered `bursty'. The exception in the right panel of Figure \ref{fig:birthrate} is the large OB associations in Bricks 15 and 21 where we know that SF is occurring at an elevated rate. These low values of $b$ are not wholly unexpected since the 3.6 \micron\ image is quite smooth (see the upper left panel in Figure \ref{fig:global_mass_map}) though one can still make out the ring feature.

Integrated over the entire survey area, we find that the birthrate parameter is 0.23 over the last 100 Myr and 0.27 over the last 400 Myr. This further shows that M31 has been forming stars in the last 400 Myr at a much lower rate than it was in the past.

\subsection{Comparison with Previous Work}

Here we examine the results presented in this paper in relation to earlier studies of the recent SFH of M31. 

\citet{Williams2002a} measured the SFH using archival \textit{HST} data in various fields throughout the galaxy, some of which overlap with the PHAT survey area. He also recovered a SFH using CMD analysis, with an earlier version of the code used in this work. The overlapping fields in his study fall primarily along the 10-kpc ring, so they probe regions of higher SFR. His `INNER' field, which falls in B13, shows a SFH with a steep decline from 1 Gyr to \about200 Myr ago, then a short rise to a peak in a time bin covering the range 40-80 Myr ago, and a decrease to the present day. Two nearby fields, G287 and G11, show a similar morphology in their SFHs, though the peak comes at slightly earlier times, closer to 100 Myr. Other fields fall along the northeast section of the 10-kpc ring and show strong SF at the most recent times. Our SFHs are consistent with this. As we showed in Figure \ref{fig:theta_sfh}, the average SFR per time bin can vary significantly in different regions. If we were to define our regions differently, such that they fell only on the regions of strongest SF, we would also see a sharper rise to the present day. 

\citet{Williams2003a} measured the SFH of M31 using ground-based data \citep{Massey2006a} and found an increase in the SFR from \about25 Myr ago to the present. This increase was seen primarily in the northeast spiral arm. While the work we present here shows a decrease in the SFR from \about50 Myr to the present, our last time bin in Figure \ref{fig:global_sfh_map} from 25 Myr to the present shows a tightening of the ring structure and more localized SF along the northeast arm, suggesting that the difference in resolution between this study and that of \citet{Williams2003a} could be the cause of the discrepancy. For example, blends could be measured as upper MS stars, artificially increasing the measurements in the youngest time bins from ground-based data.

\citet{Davidge2012a} examined the recent SFH of the entire disk of M31 by comparing $u'$ luminosity functions with those derived from models assuming various SFHs. Their result indicates a factor of 2-3 rise in the SFR during the past 10 Myr, which is in broad agreement with the results of \citet{Williams2003a}.

While our results do not show a recent SFR increase, we will point out that \citet{Williams2003a} and \citet{Davidge2012a} looked at the SFH over the entire disk of M31, while we focus only on \about1/3 of the disk. There is evidence that the SFR is elevated in the outer regions of the southern and western parts of the disk more so than in the eastern parts. Our study does not include the southern disk, nor does it reach the outer regions of the western edge of the disk. It is possible that these regions could contribute to an increase in the last 10 -- 25 Myr that we do not see in this work.

\citet{Williams2003a} also saw movement of SF across the disk, which he interpreted as evidence of propagating density waves from the northern to the southern disk. We do not see this movement. However, our survey does not cover the southern disk and our region sizes are much smaller, giving us the ability to more precisely locate SF. We find no evidence for propagation in the area covered by the PHAT survey. We do, however, agree that SF has been confined primarily to the ring structures over at least the last 500 Myr.

\mbox{~}
\subsection{The Mystery of the 10-kpc Ring}
\label{subsec:ring}

\subsubsection{The Ring is Long Lived}
The results we present here indicate that the ring of SF at a radius of 10 kpc has persisted for at least 500 Myr. If we assume a rotational velocity of M31 at the 10-kpc ring of \about250 km s$^{-1}$ \citep{Chemin2009a, Corbelli2010a}, then the dynamical time of M31 at the ring is \about250 Myr. We have defined the dynamical time as the time to make one full rotation at a given radius: $t_{\rm dyn} = 2\pi \, r / v_r$. This means that SF has continued in the ring for at least two dynamical times. \citet{Williams2003a} found that SF has occurred in the ring for the last 250 Myr, and \citet{Davidge2012a} found SF in the ring for at least 100 Myr. \citet{Dalcanton2012a} showed that there is an over density of stars with ages $>$1 Gyr in the ring as well, further supporting the long-lived nature of this feature. 

\begin{figure}[tb]
\centering
\includegraphics[width=\columnwidth]{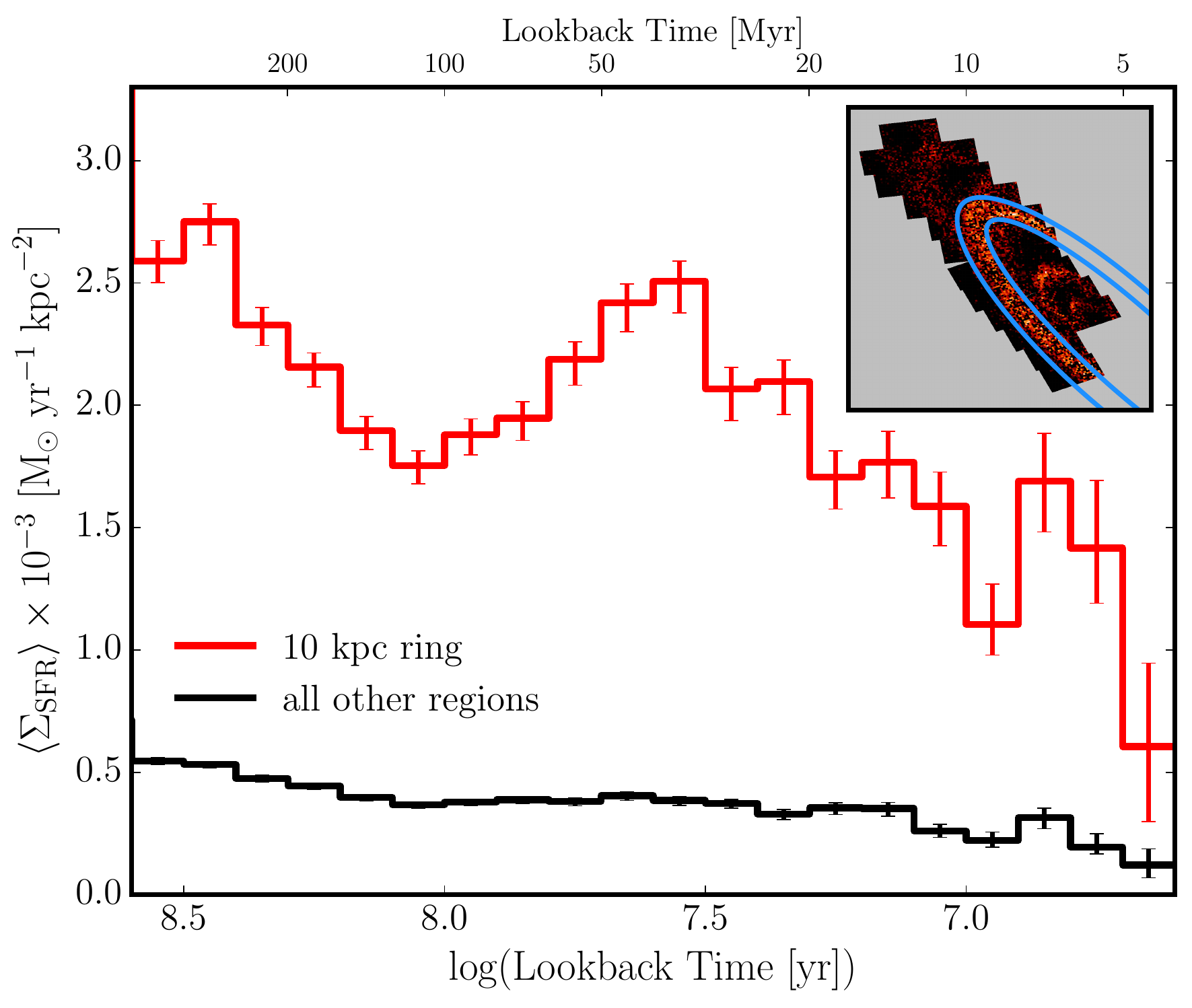}
\caption{SFH showing the average SFR in each time bin for all regions within the 10-kpc ring (red) and all other regions (black). Error bars include random uncertainties primarily due to the number of stars on the CMD as well as uncertainties as a result of the search in \av\ and \dav\ for the best-fit SFH. Comparison with Figure \ref{fig:global_sfh} shows that the total SFH is driven by SF within the ring. SF outside of the ring feature was more important before \about500 Myr ago. It is important to note that due to the high inclination of M31, our definition of the 10-kpc ring, shown in the inset, may also include some of the 15-kpc ring (if it is indeed a distinct feature). Nevertheless, this still confirms our result that SF occurs primarily in these ring features.}
\label{fig:ring_noring_sfh}
\end{figure}

Looking at just the regions that fall inside the 10-kpc ring, we see that over the last 600 Myr, SF in the ring drives the overall SFR (see Figure \ref{fig:ring_noring_sfh}). Within the last 400 Myr, the SFR in all regions that do not lie in the ring has been mostly constant, while the SFR inside the ring has varied substantially. This is also clearly visible in Figure \ref{fig:global_sfh_map}, where the ring is very prominent due to a lack of SF occurring in other parts of the galaxy. We find that in the last 400 Myr, \about60\% of all SF occurs in the 10-kpc ring feature.

The long-lived nature of the ring is also visible in Figure \ref{fig:sigma_radius}, where SF is elevated relative to the areas just outside of the ring at all time. Only in the most recent 25 Myr do we see a notable decrease in SFR in the 10-kpc ring. This plot only includes regions that fall along the major axis so as to minimize uncertainties in the deprojected distance to each region. As a result, this slice does not include many of the higher SFR OB associations, such as OB 54 and those that fall along the northeast portion of the 10-kpc ring. These would likely act to increase the SFR over the most recent \about50 Myr.

Not only is the ring long-lived, but it has also been mostly stationary. From Figure \ref{fig:sigma_radius}, we can see that the 10-kpc ring has remained centered at about the same location, moving no more than 0.5 kpc over \about500 Myr. This translates into a motion of \about1 km/s.

\subsubsection{Dispersion of the Ring}
\label{subsec:diffusion}
Our maps of SF show that the 10-kpc ring is clearly broader at older ages than at younger ages. This suggests two possibilities: 1) Over time, SF in the ring has grown more concentrated, occurring in a much narrower strip, likely following the density of gas within the ring features or 2) SF has always occurred in the center of the ring and we see a broadening of the ring as the stars disperse. The second option is more probable, since we know qualitatively that the majority of stars form in clusters and that those clusters eventually disperse and the individual stars diffuse into the surrounding environment, becoming part of the larger galactic background \citep[e.g.,][]{Harris1999a, Bastian2009a}.

The simple kinematical argument is that if stars are born with an average random velocity of 10 km s$^{-1}$ and that motion is directed radially outward, then over the course of 100 Myr, those stars should move 1 kpc. For the results we present here, this would result in significant motion over the course of 500 Myr (5 kpc). It would also suggest a strong dispersion of the ring features over relatively short periods of time, which we do not see. In reality, these timescales are much longer. \citet{Bastian2011a} showed that the evolution of the spatial structure of stellar populations in nearby dwarf galaxies occurs on many different timescales, setting \textit{lower} limits of tens to hundreds of Myr, longer than would be expected based on simple arguments. 

A detailed quantitative discussion of the evolution of spatial structure will be the subject of a future paper.

\subsubsection{What is the Origin of the Ring?}
The 10-kpc ring is the most prominent feature of M31, visible in wave bands from the UV through the radio \citep[e.g.,][]{Brinks1984a, Barmby2006a, Beaton2007a, Block2006a, Gordon2006a, Massey2006a, Nieten2006a, Gil-de-Paz2007a, Braun2009a, Chemin2009a, Fritz2012a}. Its ubiquity naturally leads to the question of its origin. Surprisingly, there have been few studies devoted to this question; its genesis remains largely uncertain.

Those studies that have attempted to answer the question have invoked one of two likely formation scenarios for a ring: (1) resonance ring due to a central bar, or (2) collisional ring due to the passage of a satellite galaxy through the disk of M31.

Bars can produce inner, outer, and/or nuclear rings in galaxies \citep{Buta1986a}. M31 shows evidence of a bar, as suggested by its boxy bulge seen in infrared imaging \citep{Beaton2007a}, though it is likely not very strong \citep{Athanassoula2006a}. The bar length has been estimated to be 4 -- 5 kpc \citep{Athanassoula2006a, Beaton2007a}. If the 10-kpc ring is due to a bar, it is unlikely to be either a nuclear ring, which generally occurs within the bar \citep[e.g.,][]{Buta1996a} or an inner ring, originating at the end of the bar \citep{Schwarz1984a}. Instead, if the ring is indeed due to a rotating bar, it must be an outer ring, which occurs near the outer Lindblad resonance \citep[OLR, e.g.,][]{Schwarz1981a, Buta1995a}. It has also been shown that in barred galaxies with rings, the ratio of the outer ring diameter to the bar diameter is \about2 \citep{Kormendy1979a, Athanassoula2009a}. By comparing observational data of M31 with N-body simulations, \citet{Athanassoula2006a} find that the OLR is at $45 \pm 4$ arcmin (9 -- 10 kpc at the distance of M31). The outer ring radius to bar length ratio and the location of the OLR in M31 suggest that the 10-kpc ring could indeed be due to a bar.

Rings in galaxies can also be the result of collisions with satellite galaxies \citep[e.g.,][]{Lynds1976a}. In the case of M31, that satellite galaxy is likely M32. A handful of studies have attempted to model the effect of M32 crashing through the disk of M31 \citep{Gordon2006a, Block2006a, Dierickx2014a} and have found that such a collision could produce a ring with properties similar to that observed, including ring centers offset from the center of the galaxy, and the hole seen in the 10-kpc ring. These studies suggest that the impact event happened 20, 210, and 800 Myr ago, respectively.

There are inconsistencies with both scenarios. It is much more statistically likely that M31 is a barred galaxy with resonance rings simply because there are many barred galaxies that show this kind of ring structure. However, the offset centers of the 5 and 10-kpc rings \citep{Block2006a} and the hole in the 10-kpc ring are difficult to explain if the rings are all due to a bar. On the other hand, the collisional simulations suffer from uncertainties in the mass and orbit of M32, both of which strongly affect the collision and resulting morphological features. In addition, if a collision occurred, we would expect to see evidence of ring expansion. We note, however, that it is possible that existing spiral structure and bar disturbances could interact strongly with these much weaker collisional rings generated by a companion, preventing the expected classical propagation \citep{Struck2010a}.

Our results indicate that the ring has been a distinct feature, actively forming stars for at least 500 Myr.  \citet{Davidge2012a} found that the last major disturbance to the disk occurred at least 500 Myr ago. \citet{Bernard2015a} and \citet{Williams2015a} both find the last major event in M31's history to be 2 -- 3 Gyr ago. These are longer than the timescales suggested by \citet{Gordon2006a} and \citet{Block2006a} for a collision with M32. We cannot rule out a collision that happened 800 Myr ago \citep{Dierickx2014a} much less 2 -- 3 Gyr ago. However, the velocity deviations suggested by collisional ring models do not match the observations presented here. \citet{Block2006a} expect the rings to expand radially at \about7 -- 10 km s$^{-1}$ and \citet{Dierickx2014a} find radial motions of at least 20 km s$^{-1}$ in their models. The systems described by \citet{Struck2010a} have radial motions of at least 10 -- 20 km s$^{-1}$, though he does note that M31 seems to be an exception to the general class of colliding ring galaxies. 

Consequently, based on existing models of the origin of M31's ring features, we can rule out a purely collisional origin because of the additional time-resolved data we present in this paper. A bar-induced ring is more likely to be long-lived and resonance rings from bars are fairly common. If the ring is a resonance ring, it would make M31 a much more average galaxy than one that had recently suffered a dramatic collision. On the other hand, perhaps the morphological features we see in M31 result from a combination of the two scenarios. We note that no existing studies attempt to explain the existence of the outer ring at 15 kpc. 

If we wish to fully understand the origin of the 10-kpc ring (or either of the other ring features) in M31, it is clear that more work is needed. One necessary step is to run simulations that follow the creation of the ring and its evolution over 1 Gyr or more. These simulations should account for multiple possible formation scenarios: collision with a satellite, resonance due to a bar, and a combination of the two. \citet{Athanassoula2006a} provide stellar velocity field predictions that need to be tested. Finally, obtaining deep photometry that goes below the oldest main sequence turnoff over a large area around the ring in combination with a more complex dust model would allow us to determine the SFH in M31 back to more than 5 Gyr ago, providing additional observational constraints on the lifetime of the ring.

\section{Conclusions}
\label{sec:conclusion}

We have measured the SFH of \about1/3 of the star-forming disk of M31 from \textit{HST} images as part of the PHAT survey. We divided the survey area into \about9000 approximately equal-sized regions and determined the SFH, as well as the extinction distribution, independently in each region. 

We have found that SF in M31 has been largely confined to three ring features, including the well-known 10-kpc ring, over the past \about500 Myr. The 15-kpc ring becomes most prominent and structured starting at about 80 Myr ago, with SF increasing to the present day. In the 5-kpc ring, SF reached a peak about 100 Myr ago and the ring feature has since begun to dissolve. 

The 10-kpc ring is long-lived and stationary, producing stars over at least the past 400 Myr, which is about two dynamical times at that radius. Over this period of time, SF has been elevated relative to the surrounding regions. As a result, the ring drives the overall SFH. The shape of the total SFH follows that of the SFH of just regions that fall broadly in the 10-kpc ring feature. This feature has also shown little or no significant propagation in our survey area over the lifetime of these SFHs.

We have shown that the total mass formed over the last 400 Myr is less than 10\% of the total mass of the galaxy, as measured from 3.6 \micron\ images. M31 is a fairly quiescent galaxy today. Aside from a handful of bright regions that coincide with OB associations, the total SFR at the present day is very low. We have shown with maps of the birthrate parameter, $b$, that only a handful of regions have $b>2$ over the last 100 Myr or 400 Myr. In most cases, the current average SFR is very low relative to the past average. Globally, the galaxy has $b=0.23$ over the last 100 Myr and $b=0.27$ over the last 400 Myr, further confirming that M31 is a fairly quiescent galaxy today.

Finally, aside from a possible increase in SF \about50 Myr ago, the SFH over the last 400 Myr has been relatively constant. We have computed the average SFR over the past 100 Myr to be $0.28\pm0.03$ \sfr\ within the PHAT footprint. Extrapolating to the entire galaxy we find a global SFR over the last 100 Myr of 0.7 \sfr, which is consistent with the range found in previous studies.

\acknowledgments

The authors thank the anonymous referee for comments that improved the clarity of this paper. We also thank
Pauline Barmby for providing us with the \textit{Spitzer} 3.6\micron\ image.
Support for this work was provided by NASA through grant number HST-GO-12055 from the Space Telescope Science Institute, which is operated by AURA, Inc., under NASA contract NAS5-26555. D.R.W. is supported by NASA through Hubble Fellowship grant HST-HF-51331.01 awarded by the Space Telescope Science Institute. This research made extensive use of NASAÕs Astrophysics Data System Bibliographic Services. The SFHs were run on the Stampede supercomputing system at TACC/UT Austin, funded by NSF award OCI-1134872. In large part, analysis and plots presented in this paper utilized iPython \citep[][]{Perez2007a} and packages from Astropy, NumPy, SciPy, and Matplotlib \citep[][]{Astropy2013a, Oliphant2007a, Hunter2007a}

\bibliography{sfh}
\mbox{~}

\clearpage

\appendix

\section{Filter Choice}
\label{app:filterchoice}

We have chosen to use only the optical filters (F475W and F814W) for this analysis. In the PHAT survey, data were also taken in two NUV filters (F275W and F336W) and two NIR filters (F110W and F160W). The optical filters provide the greatest leverage for probing the recent SFH because they have by far the deepest CMDs of MS stars which allows us to probe both further down the luminosity function and further back in lookback time. 

In principle, addition of the UV filters might potentially allow for greater constraints on the recent SFH; however, in practice, the PHAT data do not allow it. Only the brightest main sequence stars have measurements in the two UV filters, which significantly reduces the age range over which the SFH can be derived. As such, including the UV filters will not improve the analysis presented in this paper. 

The NIR data are more complex. They are significantly shallower than the optical data. The mean depth in F160W is an absolute magnitude of \about0, which is 2 magnitudes shallower than in the optical. At this depth, we lose the main sequence at less than 200 Myr. In addition, there are not very many stars that fall on the main sequence to begin with, further reducing the usefulness of the NIR for analysis of the recent SFH. Finally, the NIR models of young stars are much more uncertain than the optical models, as there has been less verification of the models in this regime.

\begin{figure*}[h]
\centering
\includegraphics[width=\textwidth]{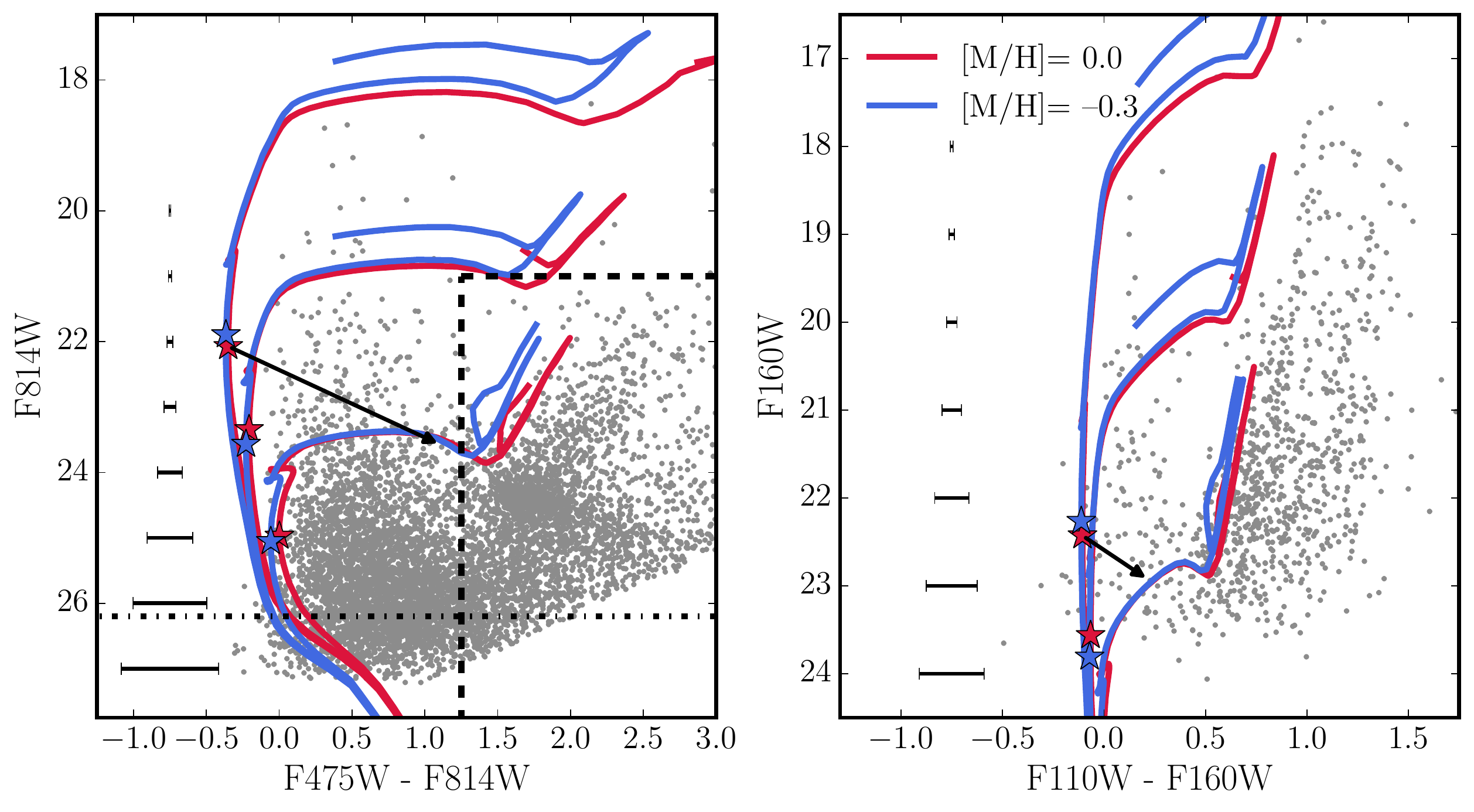}
\caption{The optical (left) and NIR (right) CMDs of a highly star-forming region in B15. On each CMD, we have plotted three sets of isochrones at 25, 100, and 400 Myr and two metallcities: solar ([M/H] = 0.0, red) and sub-solar ([M/H] = --0.3, blue). We also show a reddening vector of 2.5 magnitudes which is the amount of extinction required by the best-fit SFH in this region. In the left panel, we also show the optical exclude region described in this paper (dashed line), the 50\% completeness limit (dot-dash line).} 
\label{fig:2camiso}
\end{figure*}

In Figure \ref{fig:2camiso}, we plot the color-magnitude diagrams of a bright, highly star-forming region in B15. The left panel shows the optical data and the right panel shows the NIR data. On each panel, we have plotted three sets of isochrones at 25, 100, and 400 Myr at two different metallcities, solar and slightly sub-solar ([M/H] = --0.3). M31 has a very flat metallicity gradient that is approximately solar, so these values are reasonable for the disk. The stars mark the main sequence turnoff points at each age and metallicity. The optical data are deep, and we have leverage back at least 400 Myr. On the other hand, the IR data are much shallower and do not allow reliable SFH measurements beyond \about200 Myr. Consequently, we have chosen to use just the optical data in this analysis of the recent SFH.

\clearpage

\section{Determining the Best-Fit SFH by Searching \av, \dav\ Space}
\label{app:avdav}

The two main free parameters in the fitting process are those that describe the dust model: an extinction applied evenly to all stars, \av, and a differential extinction, \dav, that is applied in addition to \av. The differential extinction is applied as a distribution to all stars, rather than uniformly, so that each star receives some additional extinction between 0 and \dav. As a result, the total extinction applied to each star falls in the range [\av, \av\ + \dav].

\begin{figure}[h]
\includegraphics[scale=0.45, width=0.5\textwidth]{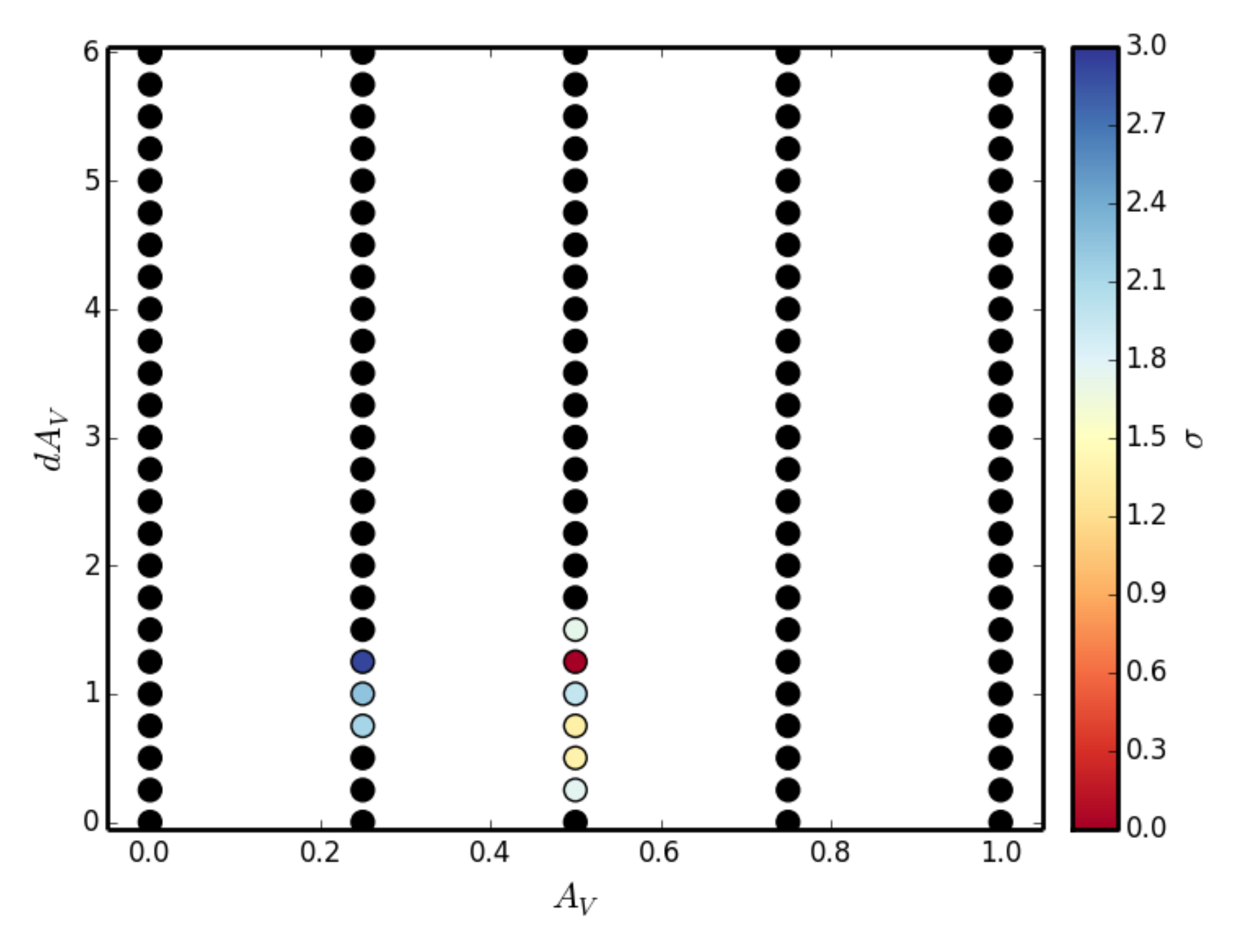}\hspace{0.01\textwidth}
\includegraphics[scale=0.45, width=0.5\textwidth]{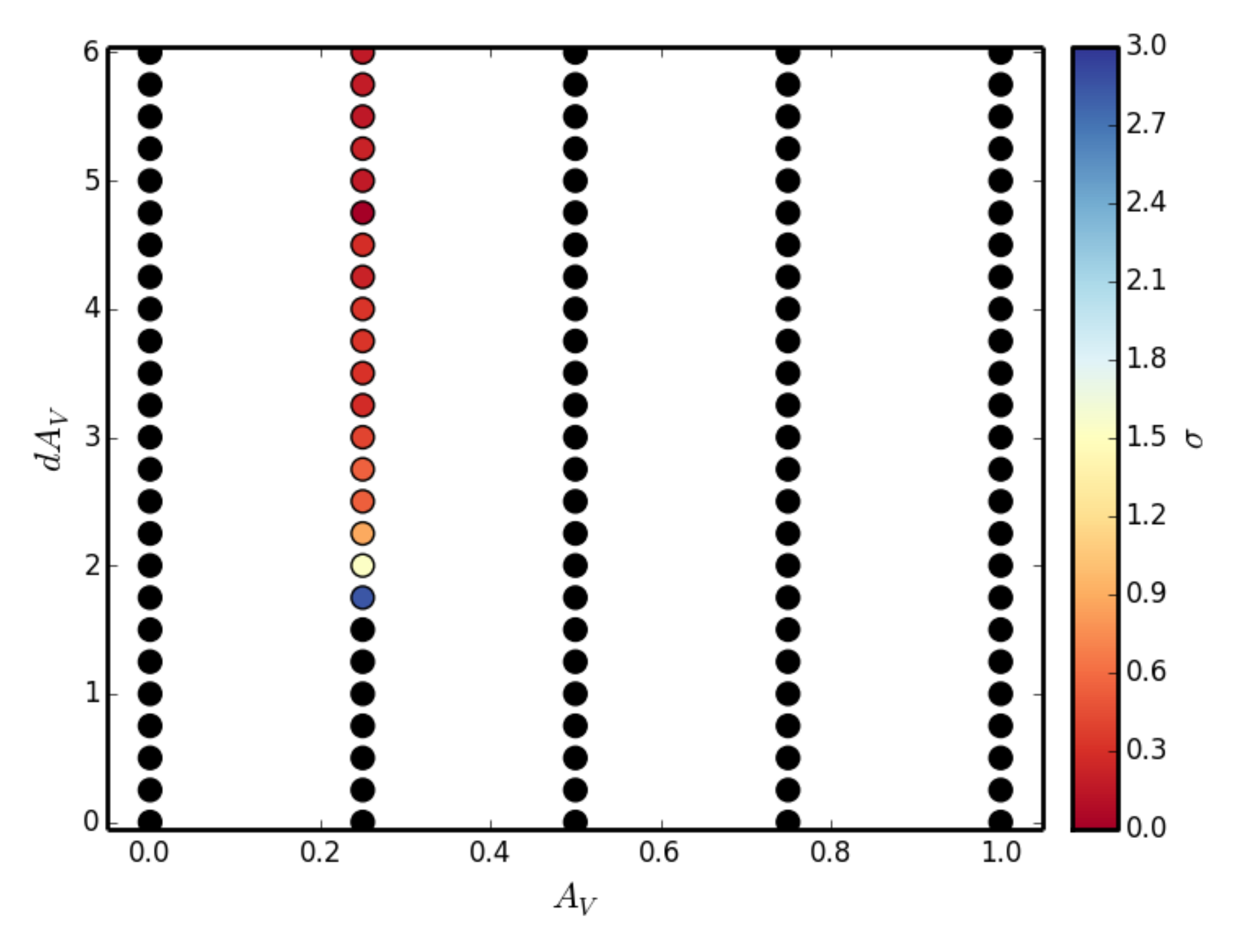}\\[0.3em]
\includegraphics[scale=0.45, width=0.5\textwidth]{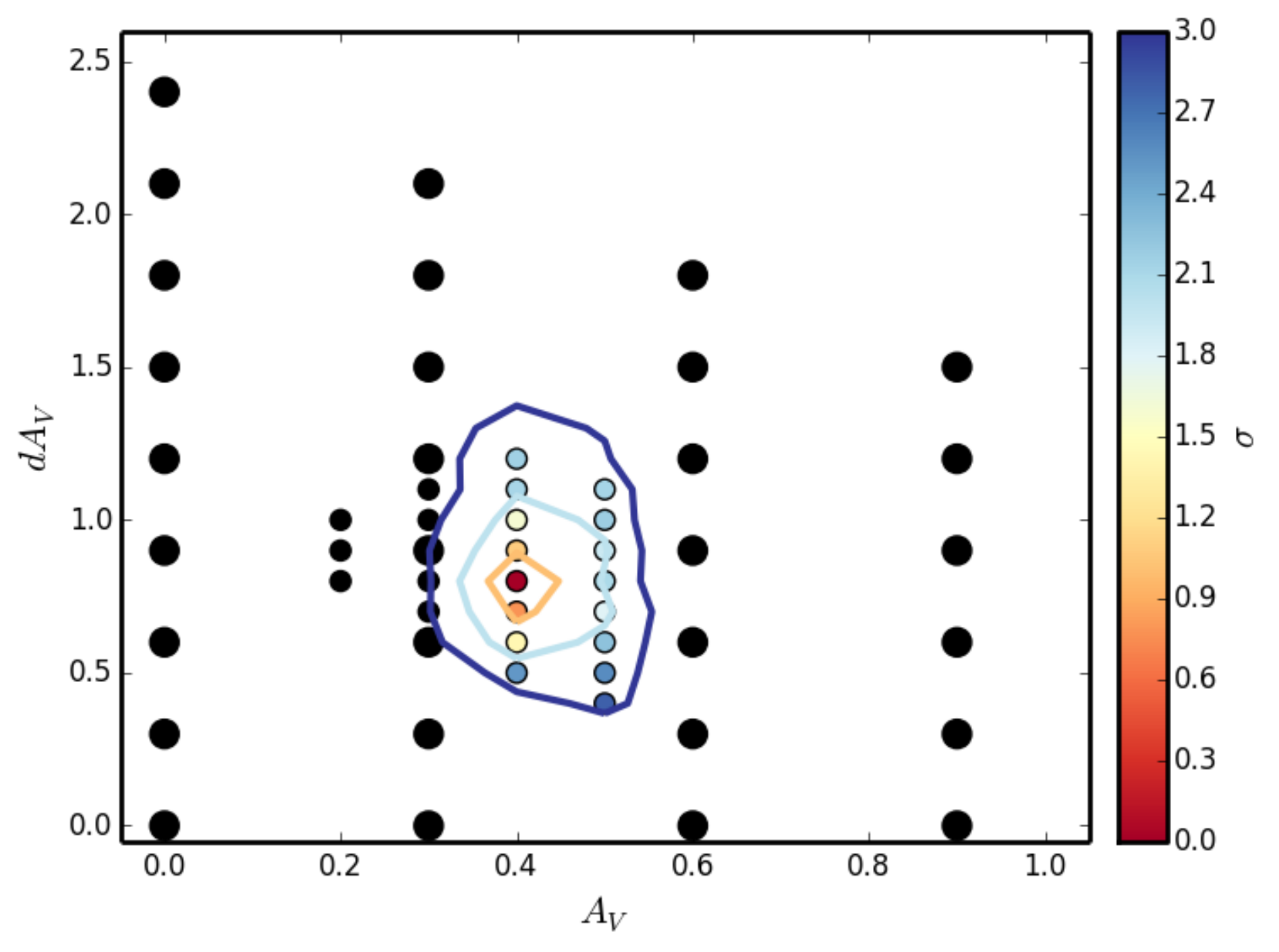}\hspace{0.01\textwidth}
\includegraphics[scale=0.45, width=0.5\textwidth]{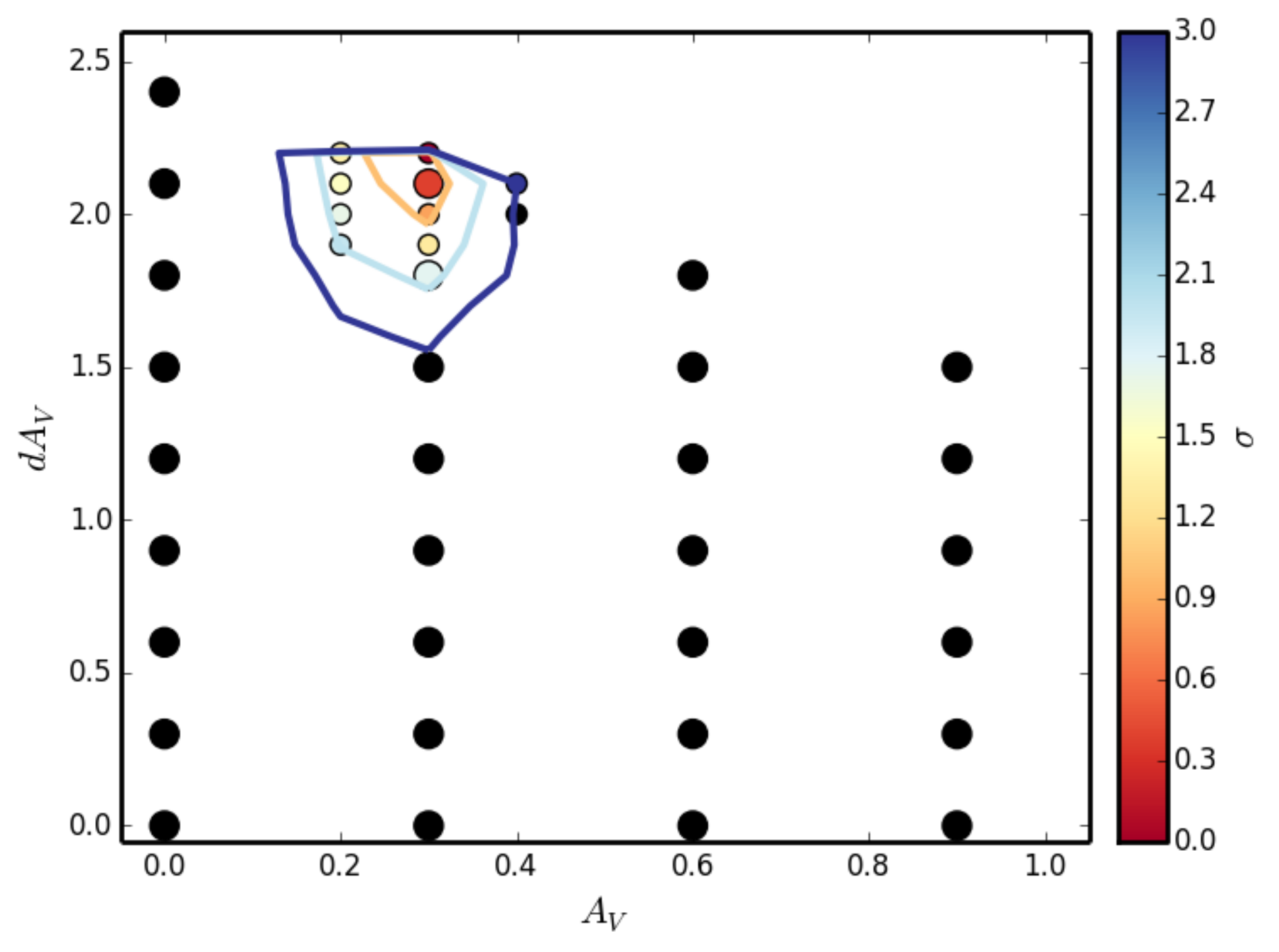}\par
\caption{Exploration of \av, \dav\ parameter space in two regions of B15. The top panel show examples of a grid search in \av, \dav\ space. One region has a very well defined best fit. In the second region, a ``trough" of low fit values occurs. If \dav\ is increased, the fit value does not change, but rather remains very close to the best fit. In the bottom panel, we show the results of the search over (\av, \dav) space for the same two regions with the requirement that \av\ + \dav\ $\le$ 2.5. Once again, the region on the left remains well-defined in \av, \dav\ space. In the second region, the constraint on extinction has forced a best fit. The lines show the 1-, 2-, and 3-$\sigma$ contours.}
\label{fig:b15_avdav_trough}
\end{figure}

Using B15, we sampled \av, \dav\ space on the range \av\ = [0.0, 1.0] and \dav\ = [0.0, 6.0] with step sizes of 0.25 in both parameters. The results for two of these regions are shown in the top panel of Figure \ref{fig:b15_avdav_trough}. We found that while some regions had a well-defined best-fit SFH, other regions showed ``troughs" of (\av, \dav) pairs going to arbitrarily high \dav\ values with fits that were within 1-$\sigma$ of the best-fit, defined as $fit_{\rm best}$ + 1. The resulting SFHs are indistinguishable within the error bars. These ``troughs" are an artifact of the exclusion area we used in the SFH solution, which removed the red clump and RGB from the fitting process. At higher extinction, stars are artificially pushed into the exclude region on the CMD. In order to account for the fewer stars in a given CMD bin, the SFR increases. Examination of the regions with these troughs shows that they tend to start at \av\ + \dav\ = 2.5. As a result, we limit our total extinction to 2.5 magnitudes.

\newpage

\section{Applying a Dust Prior}
\label{app:prior}

\begin{figure*}[t]
\centering
\includegraphics[width=\textwidth]{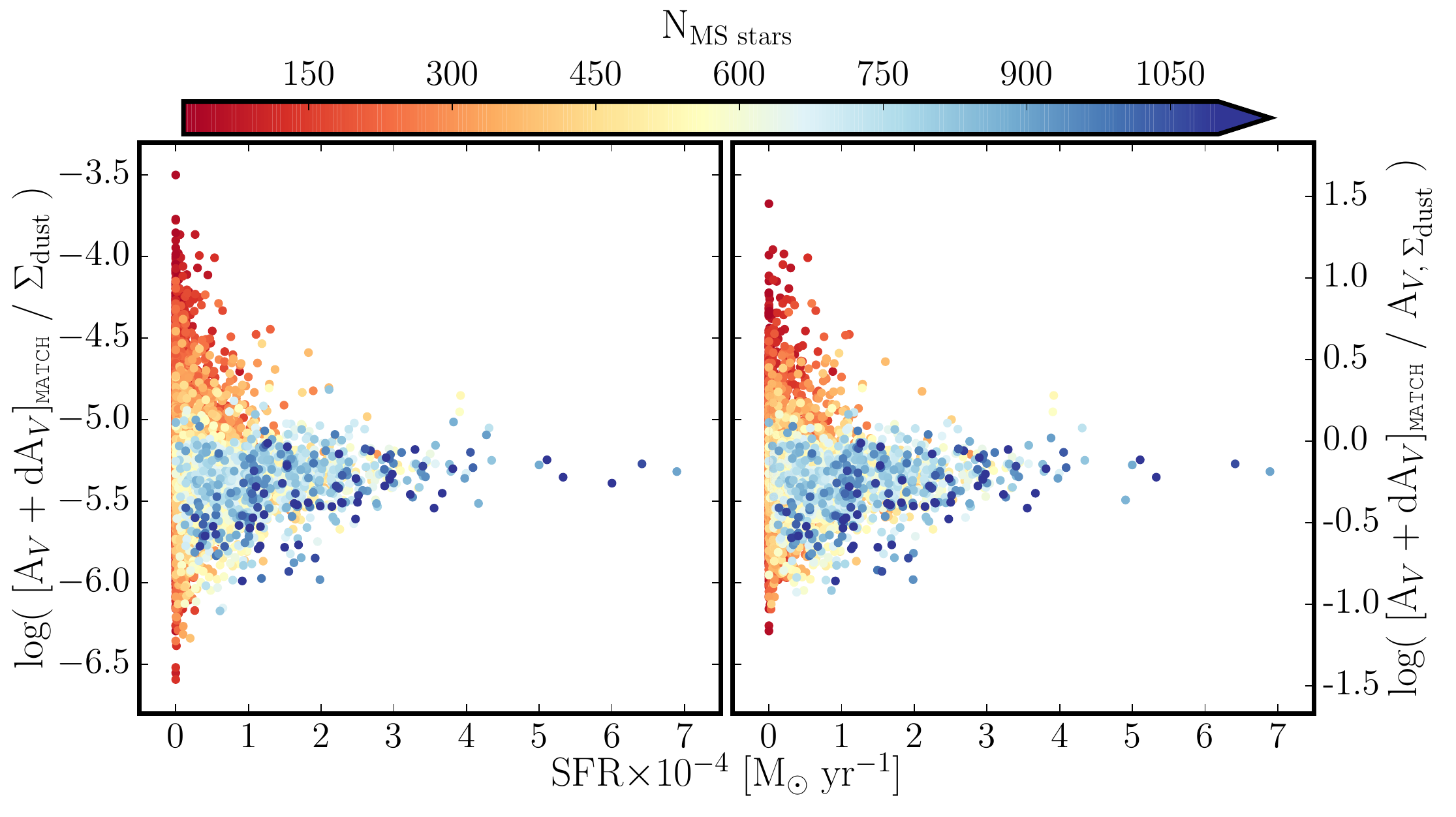}
\caption{Comparison of the \match-derived dust parameter (\av + \dav) with the dust mass surface density as derived by \citet{Draine2014a} before (left) and after (right) applying the prior. On the left axis, we show the ratio of the \match\ extinction to the dust mass surface density. On the right axis we show the ratio of the \match\ extinction to the extinction derived from the dust mass surface density following Equation \ref{eq:avconvert}. The relationship between the two parameters is generally quite constant with an expected increase in scatter at the lowest SFRs.} 
\label{fig:dust_compare}
\end{figure*}

In some of the very low-SFR regions at the survey edges and between the ring features, we found that \match\ assigned large extinctions with very poor constraints. Given the low SFR and the small number of bright, young MS stars at these locations, there is no physical explanation for large extinction. To investigate this issue, we use the \citet{Draine2014a} dust mass maps to examine the relationship between the \match\ extinction and the total dust mass in those regions. On the left side of Figure \ref{fig:dust_compare}, we plot the ratio of the \match\ total dust (\av\ + \dav) to the dust mass surface density as a function of SFR (averaged over the last 100 Myr) and number of MS stars. Each circle represents a single region, color-coded by the number of MS stars in that region, where we define MS stars as those with F475W $-$ F814W $<$ 1 and F475W $<$ 26. The bricks closer to the bulge have lower completeness, so the value of 26 was chosen to accommodate those bricks with 50\% completeness just fainter than 26th magnitude. On the right axis, we converted the dust mass surface density into extinction according to Equation 7 in \citet{Draine2014a}:
\begin{equation}
	A_V = 0.74 \left( \dfrac{\Sigma_{Md}}{10^5 \; M_\odot \; \textrm{kpc}^{-2}} \right) \textrm{mag}.
	\label{eq:avconvert}
\end{equation}

 We have made this conversion for ease of reader interpretation, though analysis of this result is beyond the scope of this paper. For the sake of this exercise, only the relative numbers are important.

If there were a perfect correlation between measured dust mass and the \match\ dust parameters, we would expect to see a straight horizontal line in this figure such that the ratio of the normalized \match\ dust to the normalized dust mass would be constant across all SFRs with some scatter expected at the low SFR end. Instead, we see that at very low SFRs there is wide scatter in this ratio, and, as may be expected, these regions also contain lower numbers of MS stars. If there are very few or no stars on the upper MS, it is more difficult to anchor the SFH. In order to fit the broadened lower MS, the amount of differential reddening is increased. Figure \ref{fig:dust_compare} shows that the ratio between \match\ dust model parameters and the dust mass is approximately constant in regions with high SFR and/or many MS stars. This implies that we should be able to fit a straight line to a plot of \match\ dust parameters vs. dust mass in these regions with the most robust fits. So we use this relation to set a prior on the total extinction in a region to adjust the low-SFR, high-dust regions to achieve more physical extinction parameters.

We compare the \match\ best-fit extinction (\av\ + \dav) with dust mass surface density ($M_{\rm dust}$) in each region. To do this, we fit a line with slope $m$ and scatter $\sigma$ to a plot of \match\ dust (\av\ + \dav) vs. $M_{\rm dust}$ of the $n$ regions with high SFR and many MS stars. 

The slope and dispersion of that line are found by solving the following equation numerically for $m$ and $\sigma$ such that $\chi^2 = 1$:

\begin{equation}
	\chi^2 = \dfrac{1}{n-1} \displaystyle \sum_n \begin{cases}
		\dfrac{ \left(\left[A_{V} + dA_{V}\right] - m \times M_{\mathrm{dust}} \right)^2} { (\overline{\sigma_{A_{V}}}_{\mathrm{reg}}^2 + \sigma^2)}, & A_V + dA_V < 2\\
		 \displaystyle \int\limits_{2}^{\infty}  \dfrac{\left( x - m \times M_{\mathrm{dust}} \right) ^2} {(\overline{\sigma_{A_{V}}}_{\mathrm{reg}}^2 + \sigma^2)} \,  \mathrm{d}x , & A_V +dA_V \ge 2
	\end{cases}
\end{equation}

where \av\ + \dav\ is that of the best-fit SFH for that region and $\overline{\sigma_{A_{V}}}_{\mathrm{reg}} = C/\sqrt{N_{\textrm{stars}}}$. $N_\textrm{stars}$ is the number of stars in the $nth$ region and $C$ is calculated such that the resulting values of $m$ and $\sigma$ do not depend on $N_\textrm{stars}$. In this case, $C=12$. We split $\chi^2$ into two solutions to account for the fact that we have set an upper limit of \av\ + \dav\ = 2.5 in our analysis. As we have shown in Appendix \ref{app:avdav}, the best-fit extinction parameters are effectively lower limits in some of our regions. As a result, we assume that all regions with \av\ + \dav\ $\ge$ 2 are lower limits.

We find that a line with $m=4.5\times10^{-6}$ and $\sigma=0.6$ results in the fit that best helps us constrain the regions with low SFR and high dust. We apply this prior to our results in all regions by recomputing the $fit$ values for each (\av, \dav) pair. The new $fit$ value is given by 

\begin{equation}
	fit_{\rm new} = fit_{\rm old} + \dfrac{ (A_V + dA_V) - m \times M_{\mathrm{dust}}} {\sigma^2}
\end{equation}

We compare the \av\ and \dav\ values corresponding to the new fits in each region with the dust mass. The results are shown on the right side of Figure \ref{fig:dust_compare}. Applying the prior did tighten up the relation a little bit, reducing the number of outliers on both tails at the low SFR end.

\end{document}